\documentclass[aps,prx,twocolumn,superscriptaddress,showpacs,floatfix]{revtex4-2}
\usepackage{comment}
\usepackage{physics}
\usepackage{amsmath,amsthm,amssymb,amsfonts, color, comment, graphicx, slashed}
\usepackage{xcolor}
\usepackage{float}
\usepackage{graphicx}
\usepackage{dcolumn}
\usepackage{bm}
\usepackage[commandnameprefix=always]{changes}
\UseRawInputEncoding 
\usepackage{dsfont}
\usepackage{hyperref}
\newcolumntype{L}{>{$}l<{$}} 

\newcommand{\bea}{\begin{eqnarray}}
\newcommand{\eea}{\end{eqnarray}}

\newcommand{\doe}{\partial}

\newcommand{\bv}{\mathbf{v}}

\newcommand{\be}{\begin{equation}}
\newcommand{\ee}{\end{equation}}
\newcommand{\bk}{{{\mathbf{k}}}}

\newcommand{\bj}{{{\mathbf{j}}}}
\newcommand{\bi}{{{\mathbf{i}}}}
\newcommand{\bx}{{{\mathbf{x}}}}
\newcommand{\bK}{{{\mathbf{K}}}}

\newcommand{\bL}{{{\mathbf{L}}}}

\newcommand{\beal}{\begin{align}}
\newcommand{\eeal}{\end{align}}

\def\eff{\mathrm{eff}}

\newcommand{\btjstrw}{\mathrel{{\rotatebox[origin=c]{90}
{$\bowtie$}}\kern-0.18em\raisebox{-.95ex}{$\bullet$}
\kern-0.5em\raisebox{.97ex}{$\bullet$}
\kern-1.12em\raisebox{.97ex}{$\bullet$}
\kern-0.52em\raisebox{-.95ex}{$\bullet$}}}

\newcommand{\btjnbrR}{{\mathrel{\rotatebox[origin=c]{90}
{$\bowtie$}}\kern-0.22em\raisebox{.9ex}{$\bullet$}
\kern-1.em\raisebox{-.8ex}{$\bullet$}}}
\newcommand{\btjnbrL}{{\mathrel{\rotatebox[origin=c]{90}
{$\bowtie$}}\kern-0.22em\raisebox{-.8ex}{$\bullet$}
\kern-1.em\raisebox{+.9ex}{$\bullet$}}}

\def\b{\beta}

\def\d{\delta}
\def\e{\epsilon}

\def\g{\gamma}

\def\t{\tau}

\def\w{\omega}

\def\vf{\varphi}
\def\bvf{\boldsymbol{\varphi}}
\def\bth{\boldsymbol{\theta}}

\def\bI{\boldsymbol{I}}

\def\bph{\boldsymbol{\phi}}

\def\mo{{\mathcal{O}}}

\begin{document}

\title{High-Sensitivity Magnetometry with Scalable Mixed SQUID Arrays}
\author{Anjishnu Bose}
\email{anjishnu.bose@adelaide.edu.au}
\affiliation{Quantum and Nano Technology Group (QuaNTeG),
School of Chemical Engineering, Adelaide University,
North Terrace Campus, Adelaide, 5005, South Australia, Australia}
\author{Jonathan Marenkovic}
\affiliation{Quantum and Nano Technology Group (QuaNTeG),
School of Chemical Engineering, Adelaide University,
North Terrace Campus, Adelaide, 5005, South Australia, Australia}
\author{Giuseppe C. Tettamanzi}
\affiliation{Quantum and Nano Technology Group (QuaNTeG),
School of Chemical Engineering, Adelaide University,
North Terrace Campus, Adelaide, 5005, South Australia, Australia}

\date{\today}
\begin{abstract}
The utilisation of Superconducting Quantum Interference Devices (SQUIDs), a loop of superconducting material hosting two Josephson junctions, has long been at the forefront of quantum sensing. To enhance performance and signal power, modern architectures connect multiple SQUIDs into two-dimensional series-parallel arrays called Superconducting Quantum Interference Filters (SQIFs). A major recent advancement, called a mixed-SQUID array, integrates junction-free ``bare" loops alongside active SQUID loops, offering a promising strategy to engineer high-performance alternatives to traditional SQIF technology. However, a rigorous theoretical optimization and performance scaling analysis for this novel geometry remains lacking. In this work, an analytical framework, validated against full numerical simulations, is developed to establish fundamental design rules for this technology. This work systematically examines how key quantum sensing parameters evolve as a function of loop inductance, parasitic coupling, and overall array dimensions. Crucially, it is demonstrated that these mixed SQUID arrays circumvent the severe inductive screening losses inherent to conventional SQIF designs, achieving a performance boost in flux sensitivity in the range of $2-4\:\times$ under realistic device parameters.
\end{abstract}

\maketitle

\section{Introduction}
Superconducting Quantum Interference Devices (SQUIDs) form the foundational basis of modern electromagnetic detection technologies utilizing superconductivity~\cite{tinkham_introduction_2015, fagaly_superconducting_2006, kleiner_superconducting_2004, seidel_superconducting_2015, koelle_high-transition-temperature_1999}. There are many possible SQUID configurations, but the most widespread is the simple DC SQUID consisting of a loop of superconductor interrupted by two non-superconducting Josephson junctions (JJ) in parallel. A combination of the Meissner effect and the quantum origins of superconductivity leads to any magnetic flux in such loops remaining trapped only in multiples of the flux quanta, \(\Phi_0 = h/2e\approx2.07\:\mathrm{fWb}\)~\cite{tinkham_introduction_2015, annett_superconductivity_2004, fossheim_superconductivity_2005, buckel_superconductivity_2004}. Due to the unique current-voltage dynamics of a Josephson junction~\cite{Likharev1986, Ambegaokar_1969, Shapiro_1963, Barone1982physics, Stewart1968current, McCumber1968effect}, a SQUID transforms this trapped flux into a voltage readout. This forms the basis of the ultra-high sensitivity that this technology can provide. \par 

Indeed, modern SQUIDs possess some of the highest known magnetic field sensitivity and near-quantum limited energy resolution compared to other conventional detectors, in the range of \(\mathrm{pT}/\sqrt{\mathrm{Hz}}\)~\cite{Drung2007highly, Schmelz_2016_ultralownoise, Vasyukov2013scanning, RevModPhys.92.021001}. As a result, SQUIDs have been implemented into a variety of practical and industrial applications~\cite{Clarke2006squidvol2} ranging from bio-magnetism and medical imaging \cite{Hamalainen1993magnetoencephalography}, scanning magnetic imaging of circuits \cite{Kirtley2010fundamental}, detecting structural defects \cite{Jenks1997squids}, geological exploration \cite{leslie_application_2008}, and Radio-Frequency (RF) amplifiers~\cite{Muck2010radio} to name a few. SQUIDs also provide one of the most promising platforms to realize quantum computing qubits \cite{krantz_quantum_2019, Devoret2004superconducting, esteve_course_2004, Clarke2008superconducting}, and are actively being developed and studied currently.
\par

Despite their performance, a fundamental drawback of standard DC SQUIDs is the strictly periodic nature of their voltage response to applied flux. Consequently, a single SQUID operates only as a relative magnetometer unless integrated with complex, and bandwidth debilitating, flux-locked loop (FLL) feedback electronics~\cite{drung_high-performance_2002, Drung2010simplified}. To achieve absolute magnetometry without feedback circuitry, modern designs employ series-parallel arrays of SQUIDs engineered with incommensurate (non-uniform) loop areas~\cite{oppenlander_superconducting_2001, Haussler2001nonperiodic,oppenlander_non-periodic_2000,  mitchell_quantum_2019}. Known as Superconducting Quantum Interference Filters (SQIFs), these devices form the core of current high-performance unshielded magnetometers and RF electronics. However, conventional SQIF designs suffer from inherent physical trade-offs: introducing a wide distribution of loop areas drastically complicates lithographic fabrication and induces severe, spatially inhomogeneous inductive screening across the array. Furthermore, since inductive screening scales aggressively with array size (both individual loop dimensions and total junction count), conventional SQIFs cannot be scaled up efficiently without incurring severe performance degradation.\par 

To overcome these limitations, a novel design was recently proposed~\cite{Monaghan_synthetic_2025, patent} that incorporates bare (junction-free) superconducting loops into a uniform SQUID array, dubbed a ``mixed SQUID array (SQA)", enabling \emph{absolute magnetometry} while preserving structural uniformity. However, a comprehensive theoretical understanding of how the effective area distribution behaves, how inductive coupling alters device dynamics, and how performance scales with array dimensions remains lacking. In this paper, we present a rigorous study of mixed SQUID arrays, elucidating the effective re-parametrization of loop areas and inductances driven by mutual inductive coupling between the bare loops and the active SQUID array. We demonstrate that mixed SQAs achieve superior flux sensitivity compared to traditional SQIF architectures. Crucially, we show that mixed SQAs resolve a major two-fold bottleneck: (a) they significantly streamline microfabrication by eliminating physical area spreads and reducing the required junction count, and (b) they actively counter performance-degrading inductive screening, unlocking favorable scaling characteristics in large-scale devices.\par

This paper is organised as follows: in Section \ref{sec: model} we describe the model used to study the SQUID array and set up all the equations. In Section \ref{sec: effective device}, we present the effect of introducing bare loops in the SQUID array on device characteristics like loop areas and inductances. In Section ~\ref{sec : low inductance}, we set up the perturbation series in a weakly-coupled limit, and obtain analytical expressions for a variety of device performance metrics like the voltage sensitivity. In Section \ref{sec: realistic inductance}, we compare the analytical approximations to numerically simulated results for more realistic inductances, and compare the properties of the mixed SQA with conventional SQIFs. Lastly, in Section \ref{sec: scaling} we look at scaling of device parameters with system sizes. \par

\section{Model}\label{sec: model}

\begin{figure}[!ht]
    \centering
    \includegraphics[width=1.0\linewidth]{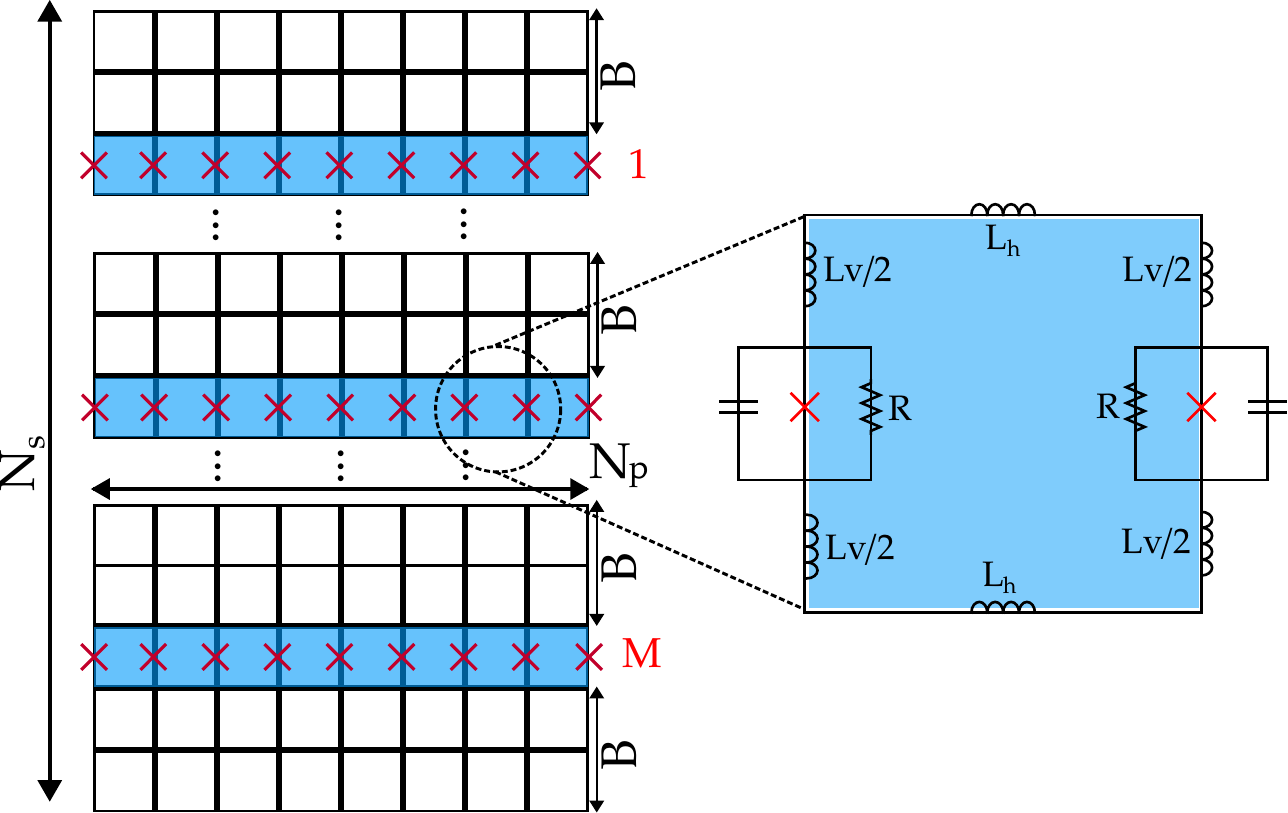}
    \caption{Schematic of a \(N_s\times N_p\) mixed SQUID array with \(B\) bare loops between each of the \(M\) full SQUID rows marked in blue with \(N_p+1\) Josephson junctions on each of the full rows marked with red crosses. (inset) A single SQUID loop consisting of two Josephson junctions with net effective resistance of \(R\), and a critical current of \(I_c\) each. The SQUID loop consists of four inductive wire segments with \(L_h\) on the horizontal wires, and \(L_v\) on the vertical wire segments.}
    \label{fig:device schematic}
\end{figure}

The response of a SQUID array can be modelled using a phenomenological circuit model called the \emph{resistive and capacitive shunted junction} (RCSJ) model \cite{Likharev1986, cawthorne_complex_1998, tesche_dc_1977, Chesca1998analytical, kornev_superconducting_2014, Berggren2012computational}. In this model, each Josephson junction (JJ) is treated as a lumped element with an effective resistance \(R\), an effective junction capacitance \(C\), and a junction critical current \(I_c\). Given gauge-invariant phases across \(N_{JJ}\) junctions \(\bvf=(\vf_1,...,\vf_{N_{JJ}})\), the total current flowing across the junctions, \(\bi = (i_1, ..., i_{N_{JJ}})\), can be decomposed into a phase-dependent supercurrent, a resistive normal current, and a displacement current as
\begin{equation}
    \label{current through JJ}
    \bi = \sin(\bvf) + \frac{\doe\bvf}{\doe \t} + \b_c \frac{\doe^2\bvf}{\doe\t^2}\,.
\end{equation}
In Eq.\eqref{current through JJ}, we have normalised the current with the junction critical current \(I_c\), the voltage with the critical voltage \(V_c = I_c R\), and time with \(t_c = \w_c^{-1} = \Phi_0/(2\pi V_c)\). We have used the Josephson relation for the voltage \(V = (\Phi_0/2\pi) d\vf/dt \), and \(\b_c=\w_c RC\) is the Stewart-McCumber parameter \cite{Stewart1968current, McCumber1968effect} which controls hysteretic behaviour. In this paper, we consider a generic \(N_s\times N_p\) mixed SQA with a mix of \(M\) rows filled with \(N_p+1\) uniform Josephson junctions in parallel, and \(B\) bare superconducting rows with no junctions between each active SQUID row, as shown in Fig.~\ref{fig:device schematic} \cite{Monaghan_synthetic_2025, patent}. The SQA is biased with a DC current \(i_b\equiv I_b/((N_p+1)I_c)\), and studied in a background of uniform external flux \(\phi\equiv\Phi/\Phi_0\) through all the bare and superconducting loops. \par

Applying flux quantization across both SQUID and bare loop sub-lattices allows us to integrate out the bare loop degrees of freedom (see Appendix \ref{app: circuit modelling} for full derivation, and explicit definitions of the inductance matrices \(L_S\,,L_C\,,L_B\,,\) and \(\bL_b\) as well as incidence matrices \(K\) and \(\bK_b\)). The resulting non-linear phase dynamics are governed by an effective active SQUID array with inductively renormalized parameters 
\begin{equation}
    \label{effective parameters}
    \begin{split}
        \boldsymbol{\phi}_{\mathrm{eff}} &= \left(\boldsymbol{\phi}_{S}-L_C\cdot L_B^{-1}\cdot\boldsymbol{\phi}_B\right)\,,\\
        L_{\mathrm{eff}} &= \left(L_{S}-L_C\cdot L_B^{-1}\cdot L_C^{T}\right)\,,\\
        \bL_{b}^{\mathrm{eff}} & = \left(\bL_{b}^{S}-L_C\cdot L_B^{-1}\cdot \bL_{b}^{B}\right)\,,\\
        \bK_{b}^{\mathrm{eff}} &= \left(\bK_{b}-K\cdot L_{\mathrm{eff}}^{-1}\cdot \bL_{b}^{\mathrm{eff}}\right)\,,
    \end{split}
\end{equation}
where the subscripts \(S\) and \(B\) represent the SQUID, and bare loops respectively. Solving for the phases \(\bvf(\t)\) and the voltages \(\bv(\t)=d\bvf/d\t\) gives us the DC voltage \(\bar{\bv}\). These are then summed over all the junctions \(n\in\{1, 2, ..., N_p+1\}\) in each of the rows \(m\in\{1, 2, ..., M\}\) in the entire device to get the total voltage across the array \(\bar{v}\) as
\begin{equation}
    \label{DC voltage}
    \bar{v} \equiv \frac{1}{N_p+1}\sum_{m=1}^{M}\sum_{n=1}^{(N_p+1)}\left(\lim_{T\rightarrow\infty}\frac{ \varphi_{(m, n)}(T)}{T}\right)\,.
\end{equation}

\section{Effective Area and Inductance Redistribution}\label{sec: effective device}

In this section, we examine how the presence of junction-free ``bare" loops alters the fundamental physical properties of the active SQUID array. Rather than physically modifying loop geometries, mutual inductive screening currents flowing through the bare loops act as magnetic flux channels, systematically redistributing magnetic flux and renormalising the self-inductance experienced by the active  \cite{Monaghan_synthetic_2025, patent}.\par

To clarify the response and the performances of mixed SQUID arrays, we first analyse how applied magnetic flux redistributes into an effective magnetic flux vector, \(\boldsymbol{\phi}_{\eff}\). Assuming a uniform external magnetic field, integrating out the bare loops allows us to recast the effective flux threading the \(n^{th}\) SQUID column as an effective loop area \(a_n^{\eff}\) (see Appendix \ref{app: circuit parameters} for details)
\begin{equation}
    \label{effective areas s=1 b>1}
    \begin{split}
      a_{n}^{\eff}(r) = & (1+B)- \frac{2}{(B+1)}\sum_{b\in \mathrm{Odd}}^{B}\bigg\{\cot^2\left(\frac{b \pi}{2(B+1)}\right)\\
      & \frac{\cosh\left(\left(\frac{N_p+1}{2}-n\right)\theta_{b}(r)\right)}{\cosh\left(\left(\frac{N_p+1}{2}\right)\theta_{b}(r)\right)}\bigg\}\,,
    \end{split}
\end{equation}
where , \(r\equiv L_h/L_v\) is the \emph{inductive aspect ratio}, quantifying the ratio of horizontal to vertical inductances, and strength of coupling between SQUID and bare loops. \(B\) is the number of bare loops between each SQUID row, and \(\theta_b(r) = \cosh^{-1}(1+2r\sin^2(b\pi/2(B+1)))\). Note that the effective areas are in units of the physical uniform area of the SQUID loops.\par 

\begin{figure}[!ht]
    \centering
    \includegraphics[width=0.8\linewidth]{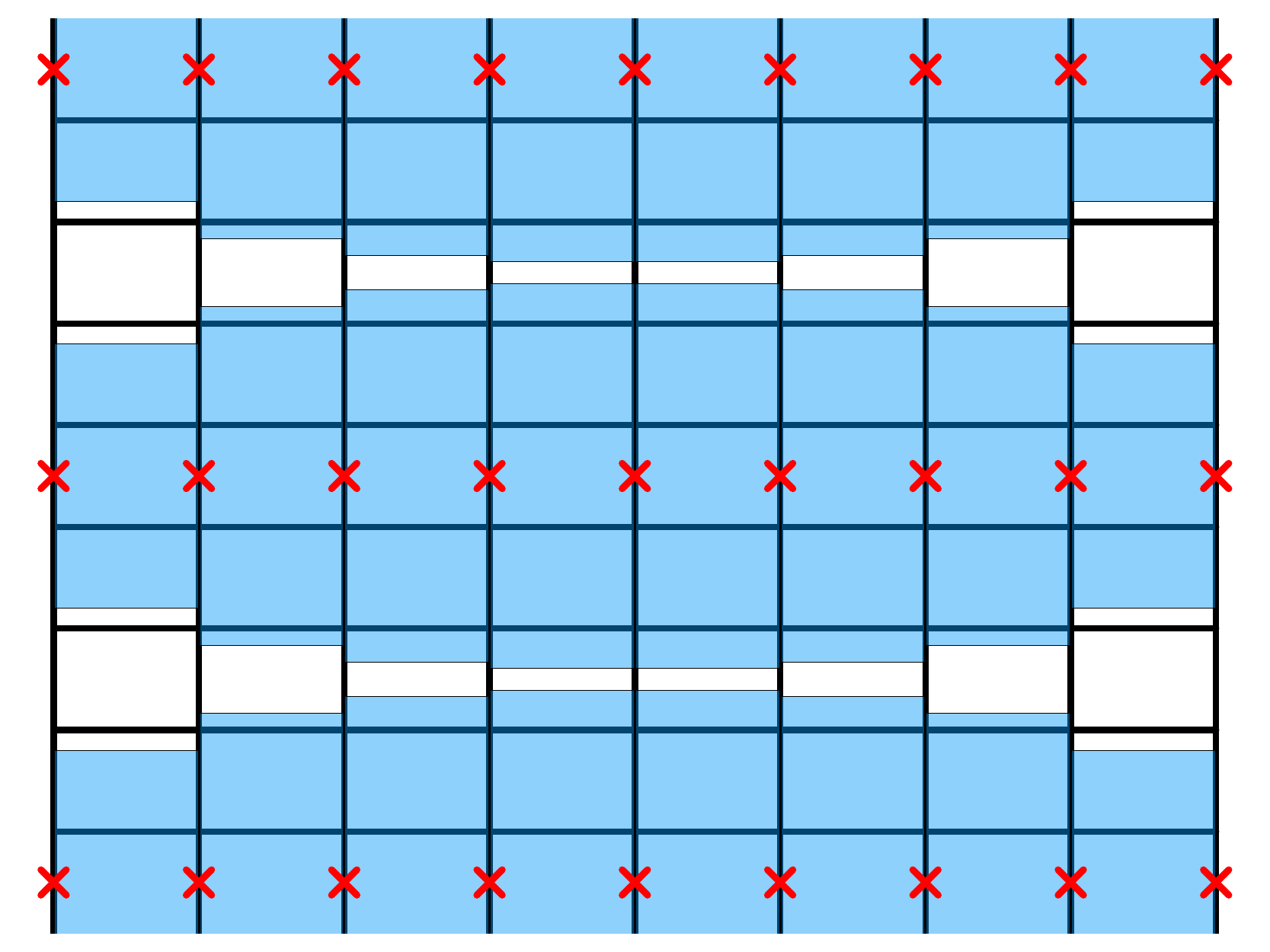}
    \caption{Typical effective area distribution for \(B=3\), \(N_p=8\), and \(r=1\). Blue area marks the effective area that the SQUID loops interact with the external flux through. The Josephson Junctions are marked using red crosses.}
    \label{fig:example area distribution}
\end{figure}

The physical behaviour of the effective areas is governed by the coupling strength \(r\). For a fixed \(B\), it varies monotonically between two limits :
\begin{itemize}
    \item \textbf{Weak coupling regime} (\(r\rightarrow 0\)) : The bare loops decouple inductively from the SQUIDs, and each loop senses only its own physical area, yielding \(a_n^{\eff}\rightarrow 1\).
    \item \textbf{Strong coupling regime} (\(r \rightarrow \infty\)) : Screening currents in the \(B\) bare loops fully focus flux into the adjacent SQUID loops. The active SQUIDs effectively absorb the flux of all neighbouring bare loops, reaching a maximum effective area of \(a_n^{\eff}\rightarrow 1+B\).
\end{itemize}

The regime of interest for practical use is the intermediate coupling (\(r \sim 1\)) range where screening currents decay exponentially from the array edges toward the center, establishing a smooth, non-uniform spatial area distribution across the columns (see Fig.~\ref{fig:example area distribution}). This generated area spread breaks phase periodicity and enables absolute magnetometry without requiring physical lithographic  \cite{Monaghan_synthetic_2025, patent}.\par

\begin{figure}[!ht]
    \centering
    \includegraphics[width=1.0\linewidth]{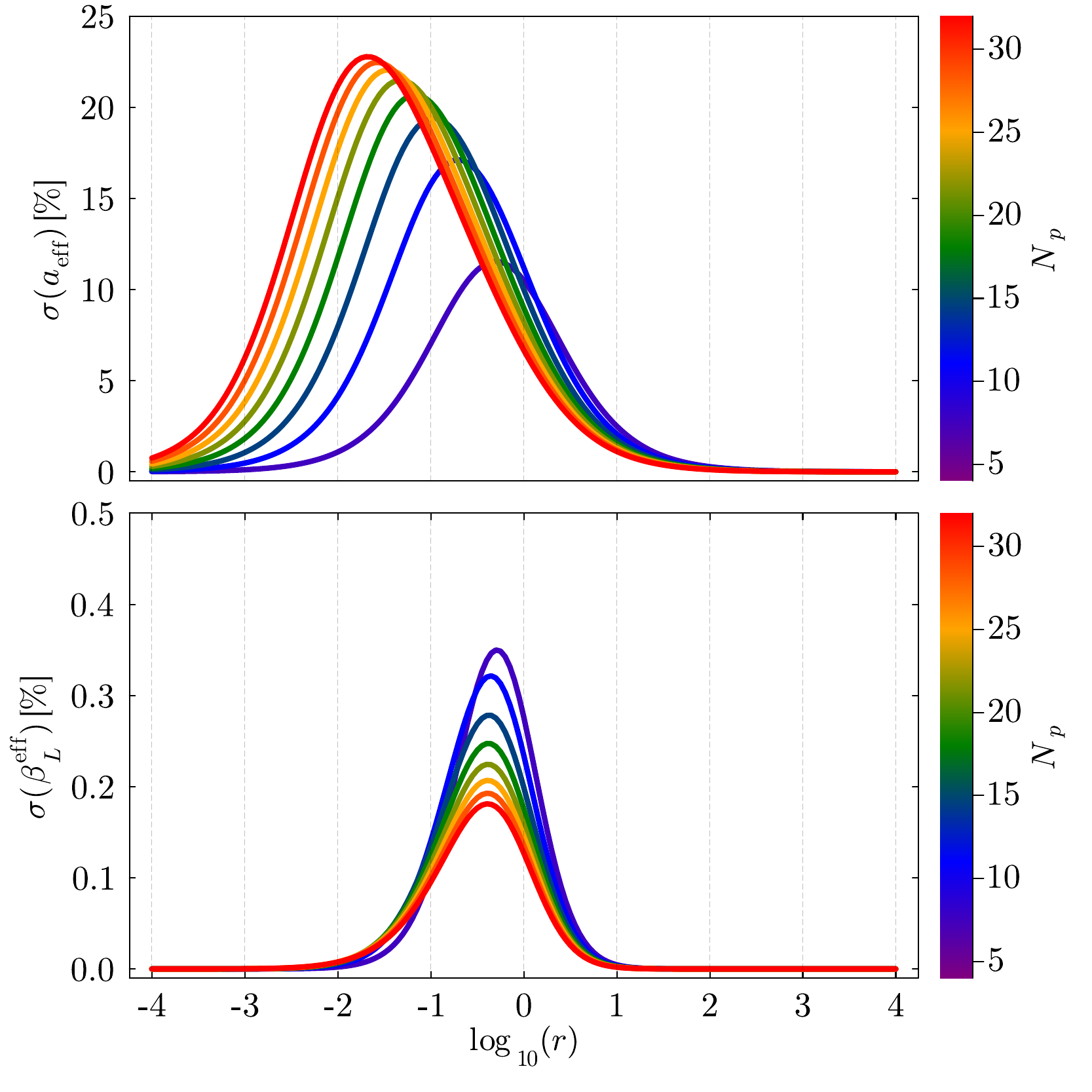}
    \caption{Comparison of the standard deviation (percentage of the average value) in the effective area distribution (top) and the effective SQUID loop self-inductance (bottom) for a device with \(B=1\) and  different sizes \(N_p=\in\{4, 8, ..., 32\}\). Note that the inductance distribution is two orders of magnitude smaller than the area distribution. A similar trend is followed for \(B > 1\) which is not shown for simplicity.}
    \label{fig:area and inductance std}
\end{figure}

To further clarify the physics in a mixed SQA, we evaluate how this inductive coupling modifies the diagonal elements of the effective inductance matrix, \(L_{\eff}\), which define the self-inductance of each SQUID loop. For a single bare loop per row (\(B=1\)), using \(L_{\eff}\) from Eq.\eqref{effective parameters}, the effective self-inductance of the \(n^{th}\) SQUID loop simplifies to:
\begin{equation}
\label{effective self-inductance}
\b_{L, n}^{\eff} = \b_L\left[1-\frac{r^2}{(1+r)}\frac{\sinh(n\theta)\sinh((N_p+1-n)\theta)}{\sinh(\theta)\sinh((N_p+1)\theta)}\right]\,,
\end{equation}
where \(\theta = \cosh^{-1}(1+r)\), and  \(\b_L = 2\pi I_c L/\Phi_0\) is the physical inductive screening parameter. Crucially, Eq.~\eqref{effective self-inductance} demonstrates that the effective self-inductance \(\b_{L, n}^{\eff}(r)\) is a monotonically decreasing function of coupling strength \(r\), asymptotically approaching \(\b_L/2\) as \(r \to \infty\). For a general array with \(B\) bare loop rows, the effective self-inductance in the strong-coupling limit is systematically suppressed to \(\b_L^{\eff} \to \b_L / (1+B)\). \par 

This reduction highlights a central advantage of mixed SQUID arrays over conventional SQIFs. In standard SQIF designs, expanding the physical loop areas to create a flux-distribution spread inherently inflates parasitic self-inductances (\(\b_{L,n}^{\mathrm{SQIF}} \propto \sqrt{\text{Area}}\)), triggering inductive screening losses \cite{kornev_design_2011, kornev_high_2013, kornev_superconducting_2014}. In contrast, as demonstrated by the standard deviations in Fig.~\ref{fig:area and inductance std}, mixed SQAs decouple area redistribution from inductance growth. The mutual inductive coupling to bare loops broadly spreads the effective area while simultaneously lowering the value and spread of effective self-inductance. This active suppression of parasitic inductance directly resolves the scaling bottleneck of conventional SQIFs (see Section~\ref{sec: realistic inductance}).

\section{Perturbative Analysis in the Low-Inductance regime}\label{sec : low inductance}

While the effective parameters derived in Section~\ref{sec: effective device} hold for arbitrary inductance, gaining analytical insight into the dynamic voltage response generally requires working in the low-inductance regime (\(\b_L \ll 1\)). In single SQUIDs operating deep in the \emph{overdamped} regime (\(\b_c\ll 1\)), perturbative expansions in powers of \(\b_L\) yield closed-form expressions for the time-dependent Josephson phases \cite{Likharev1986, soloviev_analytical_2016, berggren_analytical_2014, kolotinskiy_choice_2021, Chesca1998analytical, de_luca_pertubation_2007}. Here, this perturbative approach is extended to full 2D mixed arrays. \par 

To make the array dynamics analytically tractable, we first decompose the junction phases into two distinct physical components:
\begin{itemize}
    \item \textbf{Spatial-average row phases} (\(\theta_m\)): A set of \(M\) uniform global phases (one for each active row) that govern the secular phase growth, and time-averaged DC voltage across the array.
    \item \textbf{Phase differences} (\(\d\bvf\)): A set of \(N_p\) local phase variations across adjacent junctions within each of the \(M\) rows, which respond to local flux threading and screening currents.
\end{itemize}
Since the total current flowing through a junction row depends solely on relative phase differences, overall global phase shifts leave the branch currents invariant. This symmetry allows us to project the full non-linear RCSJ equations into a scalar differential equation for the row-average phases \(\theta_m\), and a decoupled vector system for the intra-row difference-phases \(\d\bvf\) (see Appendix~\ref{app: perturbation series} for explicit matrix projections and derivations) as
\begin{equation}
    \label{master equation decoupled scalar+vector final}
    \begin{split}
        \frac{\doe \theta_m}{\doe \tau}  & = i_b - \frac{1}{(N_p+1)}\sum_{n=1}^{N_p+1}\sin(\vf_{m, n})\,,\\
        \frac{\doe \delta\bvf}{\doe \tau} & = K^T\cdot\left[i_b\bK_b^{\eff}-\sin(\bvf)-\frac{K\cdot L_{\mathrm{eff}}^{-1}}{\b_L}\cdot\left(\delta\bvf-2\pi\boldsymbol{\phi}_{\mathrm{eff}}\right)\right] \,.
    \end{split}
\end{equation}
Expanding both phase variables in powers of the loop inductance, \(\theta_m = \theta_m^0 + \b_L \theta_m^1 + \dots\) and \(\d\bvf = \d\bvf_0 + \b_L \d\bvf_1 + \dots\), we can solve the system order-by-order to determine the fundamental performance limits of the mixed SQA.

\subsection{Leading Order Solution}

In the weak-inductance limit (\(\b_L \to 0\)), the local difference-phases settle instantaneously to the local effective flux, \(\d\bvf_0 = 2\pi \boldsymbol{\phi}_{\eff}\). At this leading order, each SQUID row is decoupled and hence the phase across every Josephson junction in an active row is driven by a \emph{single} dynamic average-phase. The spatial phase shifts generated by the effective area distribution takes the form (see Appendix \ref{app: leading order} for details)
\begin{equation}
    \label{effective phases}
    \varphi_{(m, n)}^0(\t) = 
    \begin{cases}
        \theta_m^0(\t) - 2\pi c_n(r) \phi\,,\:\:\: 1 \leq n < (N_p/2)+1\,,\\
        \theta_m^0(\t) + 2\pi c_n(r) \phi\,,\:\:\: (N_p/2) + 1 < n \leq N_p+1\,,
    \end{cases}
\end{equation}
where \(c_n(r) = \sum_{i=1}^{n}a_{(N_p/2)+i}^{\eff}(r)\) is the spatial cumulative area distribution when measured from the center. \par 

The cumulative coefficients \(c_n(r)\) govern the quantum interference dynamics of the array. In a uniform SQUID array (\(a_n = 1\)), \(c_n = n \in \mathds{Z}\) for all columns. As a result, when the external flux is an integer multiple of \(\Phi_0\), all junction phases synchronize modulo \(2\pi\), producing the strictly periodic voltage response of relative magnetometers. In a mixed SQA, however, inductive coupling to bare loops generates continuous, real-valued area coefficients (\(c_n(r) \in \mathds{R}\)). These incommensurate phase offsets prevent complete constructive interference away from zero field, suppressing secondary voltage peaks and yielding a non-periodic response suitable for absolute magnetometry \cite{oppenlander_non-periodic_2000, oppenlander_superconducting_2001, miller_enhanced_1991}. Using the area distribution in Eq.~\eqref{effective areas s=1 b>1}, \(c_n(r)\) is given explicitly by
\begin{equation}
    \label{cumulative areas s=1 b>1}
    \begin{split}
      c_{n}(r) = & (1+B)n-\frac{1}{(1+B)}\sum_{b\in\mathrm{Odd}}^{B}\bigg\{\cot^2\left(\frac{b\pi}{2(B+1)}\right)\\&\frac{\sinh(n\theta_{b})}{\sinh\left(\frac{\theta_{b}}{2}\right)\cosh\left(\left(\frac{N_p+1}{2}\right)\theta_{b}\right)}\bigg\}\,,  
    \end{split}
\end{equation}
where \(c_n(r)\) increases monotonically with coupling aspect ratio \(r\), bounded between \(n \leq c_n(r) \leq (1+B)n\).  \par

Summing the phase contributions across a row yields an effective, flux-dependent critical current for the array
\begin{equation}
    \label{modified critical current}
    i_c(\phi, r) = \frac{1+\sum_{n=1}^{N_p}\cos\left(2\pi c_n(r) \phi\right)}{N_p+1}\,,
\end{equation}
with \(i_c(\phi=0, r)=1\) always. Time-averaging the junction phase evolution under a DC bias \(i_b\) gives the total DC voltage across the \(M\)-row array \cite{Likharev1986, berggren_analytical_2014, soloviev_analytical_2016, oppenlander_non-periodic_2000, oppenlander_superconducting_2001}
\begin{equation}
    \label{DC average voltage}
    \Bar{v}(\phi, r, i_b) = \begin{cases}
        M \sqrt{i_b^2 - i_c^2(\phi, r)}\,,\:\:\: i_b\geq i_c(\phi)\,,\\
        0\,,\:\:\:i_b<i_c(\phi)\,.
    \end{cases}
\end{equation}

\begin{figure}[!ht]
    \centering
    \includegraphics[width=1.0\linewidth]{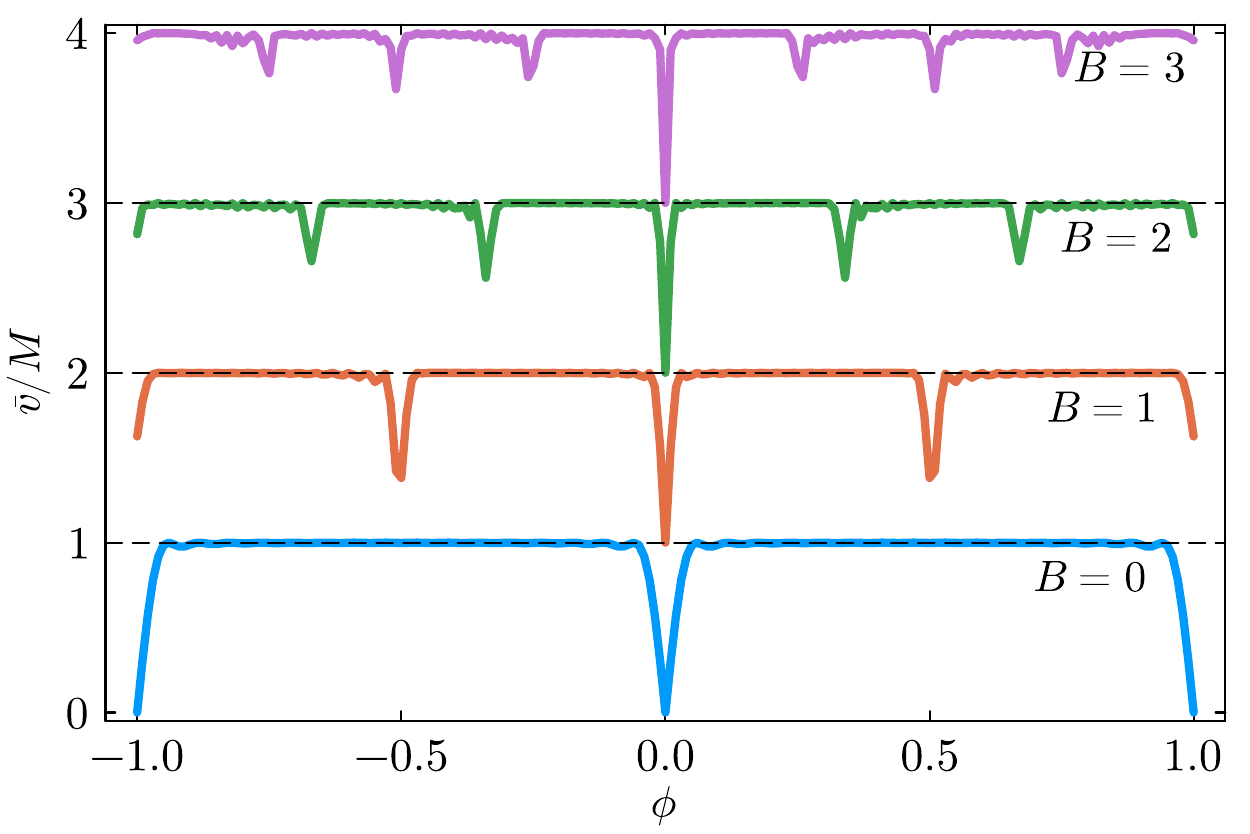}
    \caption{Comparison of the analytical \(\Bar{v}(\phi, r)\) given in Eq.\eqref{DC average voltage} at \(r=1\) for a device with \(N_p=16\), and varying values of \(B\). We have fixed the applied bias current to \(i_b=1\). We have also plotted the \(B=0\) periodic SQA with no area distribution for reference. We see that increasing \(B\) makes the voltage anti-peaks sharper, and each anti-peak appears at flux multiples of \(1/(B+1)\). }
    \label{fig:analytical vs numerical voltage flux}
\end{figure}

\begin{figure}[!ht]
    \centering
    \includegraphics[width=1.0\linewidth]{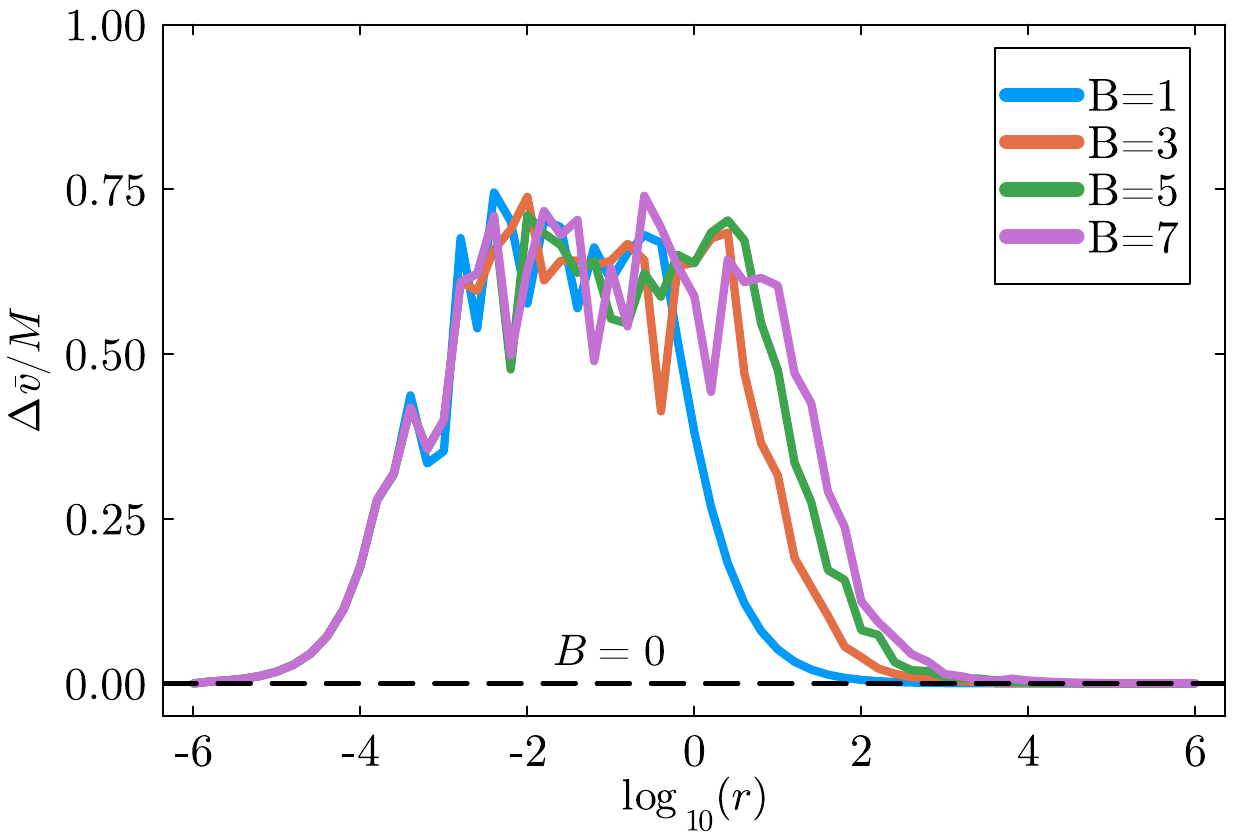}
    \caption{The voltage working range \(\Delta\Bar{v}(r)/M\) calculated using Eq.\eqref{DC average voltage} as a function of \(r\) with \(N_p=16\) and \(B=1, 3, 5\) and \(7\). The working range was numerically calculated over a range of \(\phi\in(-1000, 1000)\). Curiously the working range plateaus near \(\approx 0.75\) for all values of \(B\).}
    \label{fig: working range}
\end{figure}

As shown in Fig.~\ref{fig:analytical vs numerical voltage flux}, the resulting voltage response exhibits sharp central anti-peaks at zero field, flanked by secondary spurs that repeat at flux intervals of \(1/(B+1)\). The non-periodicity is quantified by the dynamic working range \(\Delta\Bar{v} \equiv \min_{\phi \neq 0} \Bar{v}(\phi) - \Bar{v}(0)\), which quantifies the difference between the central voltage anti-peak and the largest spur. We plot this measure in Fig.~\ref{fig: working range} and find that it correlates strongly (\(\approx 90\%\)) with the standard deviation of the fractional parts of $\{c_n(r)\}$.

\begin{figure}[!ht]
    \centering
    \includegraphics[width=1.0\linewidth]{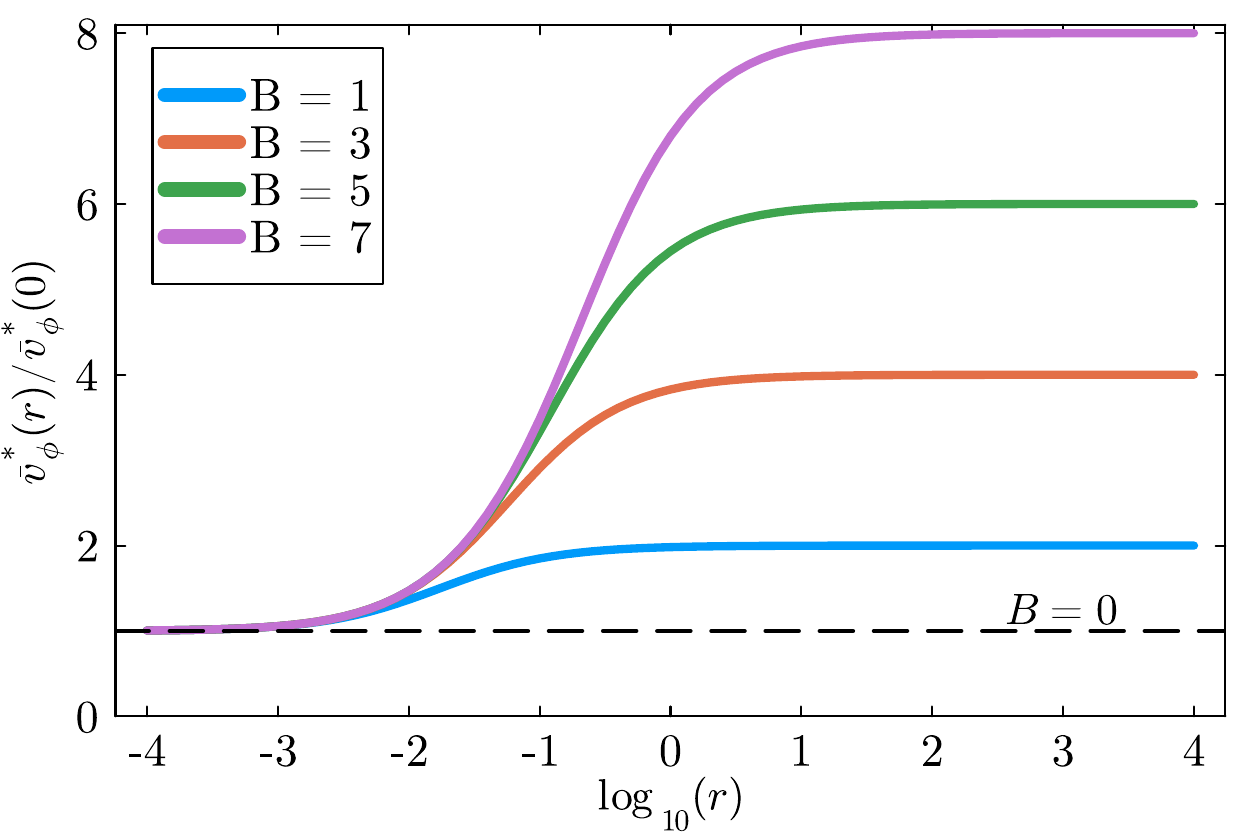}
    \caption{Maximum flux-to-voltage sensitivity calculated using Eq.\eqref{sensitivity ib=1 phi=0} as a function of \(r\) for a device with \(N_p=16\), at different values of \(B\).}
    \label{fig:voltage sensitivity}
\end{figure}

Differentiating Eq.~\eqref{DC average voltage} with respect to applied flux gives the flux-to-voltage transfer function \(\Bar{v}_\phi \equiv \doe \Bar{v} / \doe \phi\). For the device geometry studied in this paper, the maximum flux sensitivity occurs at zero applied field (\(\phi = 0\)) and critical bias current (\(i_b = 1\)) as 
\begin{equation}
    \label{sensitivity ib=1 phi=0}
    \Bar{v}_{\phi}^{*}(r) = \Bar{v}_{\phi}(\phi=0, r, i_b=1) = 2\pi M\sqrt{\frac{2\sum_{n=1}^{N_p/2}c_n^2(r)}{N_p+1}}\,.
\end{equation}
Using the bounds on \(c_n(r)\), we find that \(1 \leq \Bar{v}_{\phi}^{*}(r) / \Bar{v}_{\phi}^{*}(0) \leq B+1\) where \(\Bar{v}_{\phi}^{*}(0) = \pi M \sqrt{N_p(N_p+2)/3}\) is the maximum possible device sensitivity of a corresponding \(M\times N_p\) SQUID array \emph{without} any bare loops. In effect, incorporating \(B\) bare loops achieves a threefold advantage: it substantially increases sensitivity compared to a conventional SQUID array with the same active junction count (\(M \times N_p\)), retains the sensitivity of a fully junctioned device of equivalent physical dimensions (\(N_s \times N_p\)), and accomplishes both while successfully transforming the array into a non-periodic absolute magnetometer. We plot this leading order transfer function as a function of \(r\) in Fig.~\ref{fig:voltage sensitivity}.

\begin{figure}[!ht]
    \centering
    \includegraphics[width=1.0\linewidth]{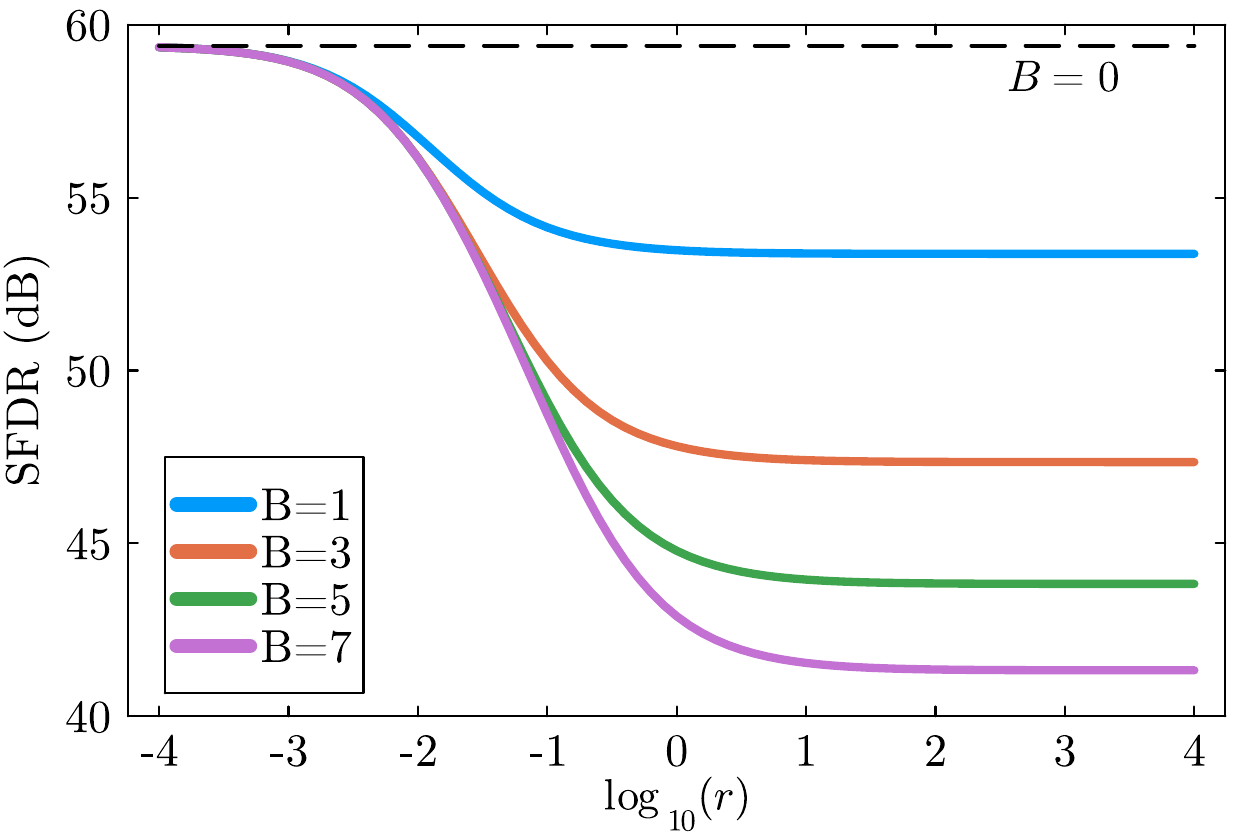}
    \caption{The spurious-free dynamic range as a function of \(r\) for a SQA with \(N_p=16\) and various fixed values of \(B\), at a critical bias current \(i_b=1\). Here, we have fixed \(\d\phi=0.01\), and  \(\w/\w_c=0.01\). }
    \label{fig:device linearity}
\end{figure}

Another important performance aspect relevant to our technology is the spurious free dynamical range (SFDR) which quantifies signal-to-output linearity. To calculate it, we first expand the voltage response around \(\phi_0 = 0\) for a small AC signal \(\phi(\t) = \d\phi \sin(\w\t)\). The ratio of the fundamental voltage response to the next highest harmonic (third-order at \(\phi=0\)) determines the (SFDR) \cite{kornev_high_2013, Kornev_2009, Berggren_2013_BiSQUID}
\begin{equation}
    \label{voltage exapansion}
    \mathrm{SFDR} =  \frac{1}{\d\phi^2}\left[\frac{(2\pi)^2}{24}\left(3 S_2+\frac{S_4}{S_2}\right)\right]^{-1}\,,
\end{equation}
where \(S_n\equiv S_n(r) = 2\sum_{i=1}^{N_p/2}c_i^n(r)/(N_p+1)\). As plotted in Fig.~\ref{fig:device linearity}, while increasing \(B\) sharpens sensitivity, it trades off slightly against SFDR, providing a clear design parameter for balancing signal gain against linearity. 

\subsection{Sub-leading Order Corrections}\label{sec: leading order correction}

At first order in loop inductance (\(\b_L\)), inductive screening currents begin to modify the phase evolution. Higher-order corrections to intra-row difference-phases (\(\d\bvf_k\)) are determined purely algebraically by lower-order phases. Consequently, only scalar differential equations are required to solve for the sub-leading average-row phases $\theta_i^1(\t)$ (see Appendix~\ref{app: sub-leading} for detailed matrix derivations). Crucially, an analytical evaluation of the first-order corrections reveals that both \(\d\bvf_1(\t)\) and \(\theta_i^1(\t)\) oscillate purely periodically in time during over-biased operation (\(i_b \geq i_c(\phi)\)). Because the time derivative of any strictly periodic function averages to zero over a full Josephson oscillation cycle, these first-order phase fluctuations contribute zero correction to the net DC voltage of the array (\(\Bar{v}_1 = 0\)).\par   

This result is fundamentally enforced by parity symmetry: under a spatial mirror reflection, the inductive circulating currents flip direction while the DC terminal voltage remains invariant. As a consequence, the DC voltage response can only contain even-power contributions in loop inductance (\(\b_L^2, \b_L^4, \dots\)). While second-order corrections (\(\b_L^2\)) become analytically intractable due to non-linear phase mixing, this symmetry guarantees that the leading-order analytical approximations derived in the previous section remain robust across small to moderate inductance values.

\section{Operational Device Performance}\label{sec: realistic inductance}

\begin{figure*}[!ht]
    \centering
     \includegraphics[width=0.475\linewidth]{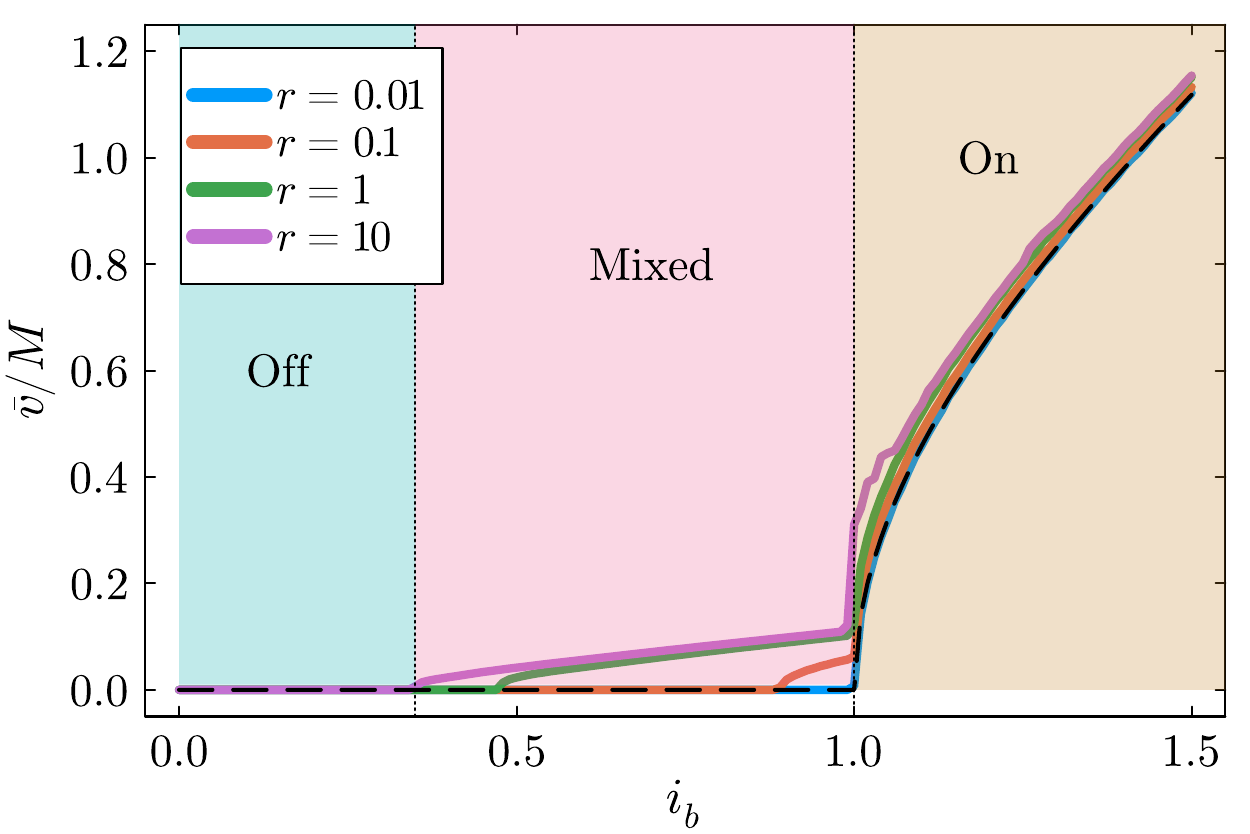}
    \includegraphics[width=0.475\linewidth]{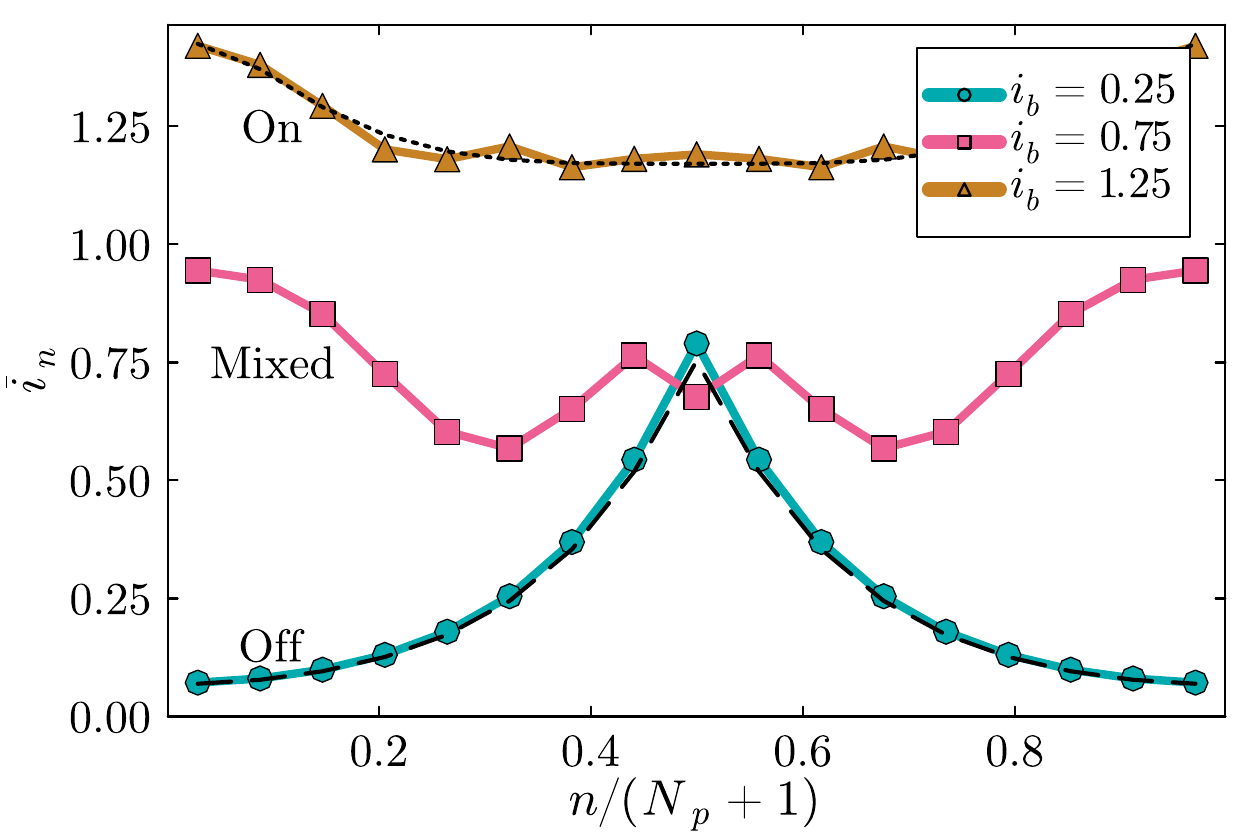}
    \caption{(Left) DC average voltage of the SQUID array \(\Bar{v}/M\) as a function of the applied bias current \(i_b\). The colored curves are from numerical simulation of the full model with \(\b_L=1.0\), \(N_p=16\), and varying values of \(r\). There are three distinct regimes of junctions being all in a off (superconducting) state, on (resistive) state, and an intermediate hybrid state. (Right) The DC branch current distribution among the junction columns for the same device at fixed aspect ratio \(r=1\) for different values of the applied bias current \(i_b\) picked from the three regimes found in the IV curve. The black dashed line is the analytical expression in the underbiased regime (see Appendix \ref{app: linearized}), while the black dotted line is the phenomenological expression for current distribution given in \cite{Berggren_2023, berggren_modeling_2024} with \(m_6\approx 0.2\).}
    \label{fig:bias current distribution}
\end{figure*}

In realistic operational regimes with finite loop inductances (\(\b_L \sim \mo(1)\)), screening currents significantly alter both the DC current distribution and the voltage response of the array. Hence, to evaluate device performance under practical conditions, a full numerical simulations of the RCSJ equations benchmarked against our results against the low-inductance analytical limits (\(\b_L \to 0\)) is performed. Furthermore, the performance of mixed SQAs are compared to conventional SQIFs featuring physical area (and proportional self-inductance) dispersion, which highlights significant performance gains and superior scaling with inductive strength.\par 

Finite loop inductance alters the device dynamics primarily by inducing current crowding near the bias injection points. For central bias injection, inductive screening hinders uniform current distribution to outer junction columns. As the applied DC bias \(i_b\) increases, this inhomogeneous current distribution drives the device through three distinct operational regimes:
\begin{itemize}
    \item \textbf{Off State} (\(i_b < i_b^*\)): 
    All Josephson junctions remain in the zero-voltage superconducting state. In this regime, linearizing the RCSJ equations yields a closed-form steady-state phase distribution \(\bvf^*\) valid to all orders in \(\b_L\) (see Appendix~\ref{app: linearized}). The branch current distribution \(\bi = \sin(\bvf^*) \approx \bvf^*\) peaks near the central injection column and decays toward the array edges. 
    \item \textbf{Hybrid State} (\(i_b^{*}\leq i_b < 1\)): 
    Inductive current crowding causes central junctions to reach their critical current (\(\varphi_{n}^* \geq \pi/2\)) and switch into a resistive state while outer junctions remain superconducting. Once these central junctions become normal, the bias current dynamically redistributes outward toward the low-impedance superconducting edges to bypass the dissipative central pathways. This causes the spatial current profile to invert, shifting from a center-peaked distribution to an edge-dominated profile. This co-existing normal and superconducting phase distribution breaks the global phase synchronization assumed in Section~\ref{sec : low inductance}, creating a localized voltage drop before the full array turns on. 
    \item \textbf{On State}: All junctions are over-biased into the resistive state, generating a dynamic time-averaged voltage response across the entire array. The dynamic branch current non-uniformity across junction columns follows a power-law profile characteristic of thin-film strip arrays with a decay exponent of \(m_6 \approx 0.2\)  \cite{Morooka1999supercurrent, Brandt1993current, Berggren_2023, berggren_modeling_2024}. Notably, this distribution emerges dynamically, and is not enforced on the model as in \cite{berggren_effects_2023}. Furthermore, current distribution becomes progressively more uniform in the inner bulk rows compared to the boundary rows \cite{gali_labarias_effect_2023}.
    
\end{itemize}

The three regimes and their corresponding current are illustrated in Fig.~\ref{fig:bias current distribution}. Evaluating the maximum steady-state phase allows us to predict the threshold bias current \(i_b^*\) marking the Off-to-Mixed transition. Importantly, this intermediate hybrid state is an artefact of localized central current injection. For homogeneous bias injection, inductive current crowding disappears, eliminating the hybrid state entirely and allowing the array to switch cleanly into the On state at $i_b = 1$. 

\begin{figure}[!ht]
    \centering
    \includegraphics[width=1.0\linewidth]{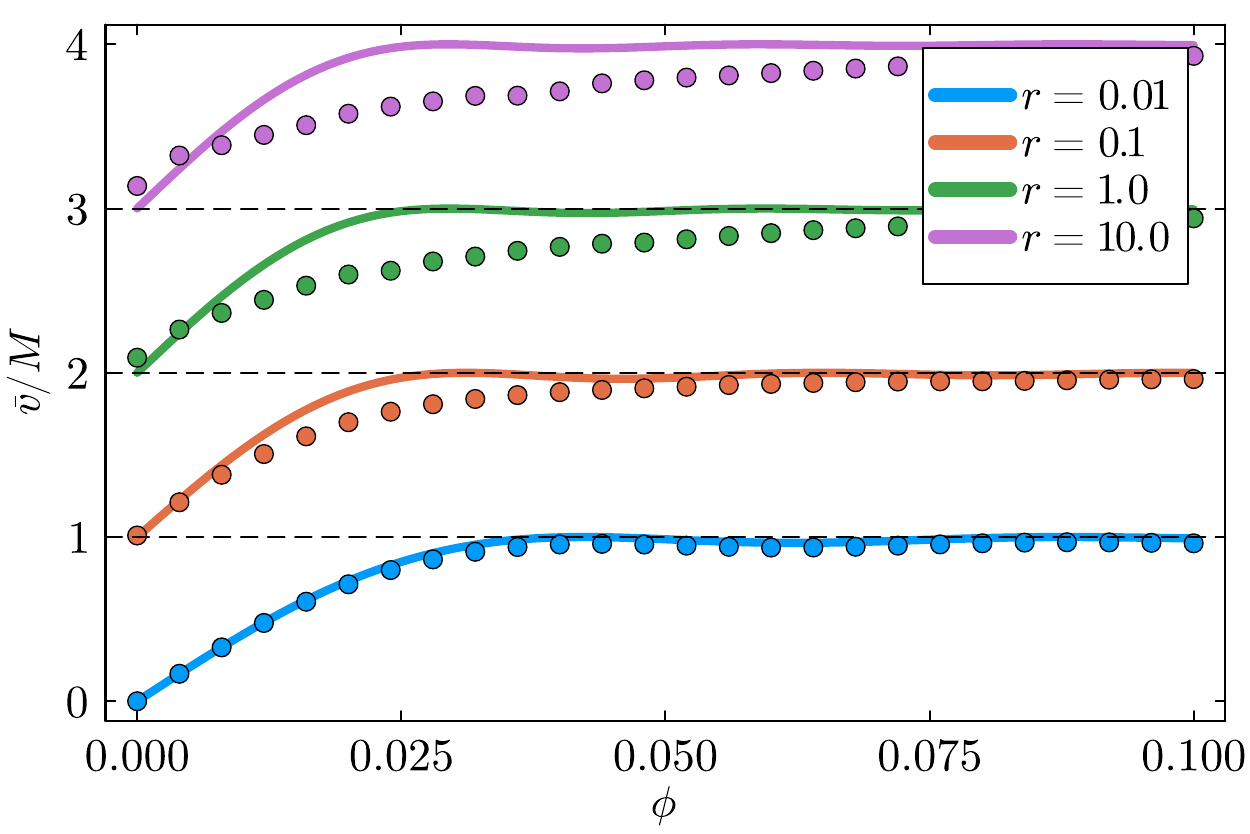}
    \caption{Comparison of the analytical \(\Bar{v}(\phi, r)\) given in Eq.\eqref{DC average voltage} for different values of \(r\) with the numerically calculated voltage at \(\b_L=1.0\), at small values of the applied flux, and a critical bias current \(i_b=1\). The solid lines mark the analytical formula while the dots represent the numerically calculated results. \(N_p=16\) and \(B=1\) were fixed in this calculation. The voltages at consecutive values of \(r\) have been shifted up for clarity. The agreement between numerics and analytics is excellent for smaller values of \(r\b_L \lesssim 0.1\), but still reasonable at realistic values of the inductance.}
    \label{fig:voltage flux realistic inductance}
\end{figure}

The characteristic voltage-flux curve of the device, and the comparison between numerical simulation and analytical expressions given in Eq.\eqref{DC average voltage} are plotted in Fig.~\ref{fig:voltage flux realistic inductance}. We find that the voltage peak-to-peak is lowered than the expected value in the presence of significant inductance. This happens simply because a large inductance leads to significant loop currents which do not contribute towards the junction resistive current and hence lower the peak voltage of the device.\par 

Next, we evaluate the maximum flux-to-voltage sensitivity of the mixed SQA as a function of finite loop inductance \(\b_L\neq 0\) and benchmark it against a conventional SQIF. For the conventional SQIF, we assume the exact same effective area distribution as the mixed SQA, but with loop inductances scaled proportionally to the physical area spread : \(\b_{L, n}^{\mathrm{SQIF}} = \b_L \sqrt{a_{n, \eff}/a_0}\). Across both architectures, we find that the maximum sensitivity generically follows a power-law scaling of the form
\begin{equation}
    \label{max sens scaling}
    \bar{v}_{\phi}^{*}(\b_L, r, B, N_p) \approx \frac{\bar{v}_{\phi}^{*}(\b_L=0, r, B, N_p)}{(1+\b_L)^{\nu(r, B, N_p)}}\,,
\end{equation}
where \(\bar{v}_{\phi}^{*}(\b_L=0, r, B, N_p)\) is given by Eq.\eqref{sensitivity ib=1 phi=0}. The scaling exponent \(\nu\) depends on the inductive aspect ratio \(r\), the number of bare loops per SQUID row \(B\), and generically on the number of columns in the array \(N_p\) as well.\par

\begin{figure*}[ht]
    \centering
    \includegraphics[width=0.475\linewidth]{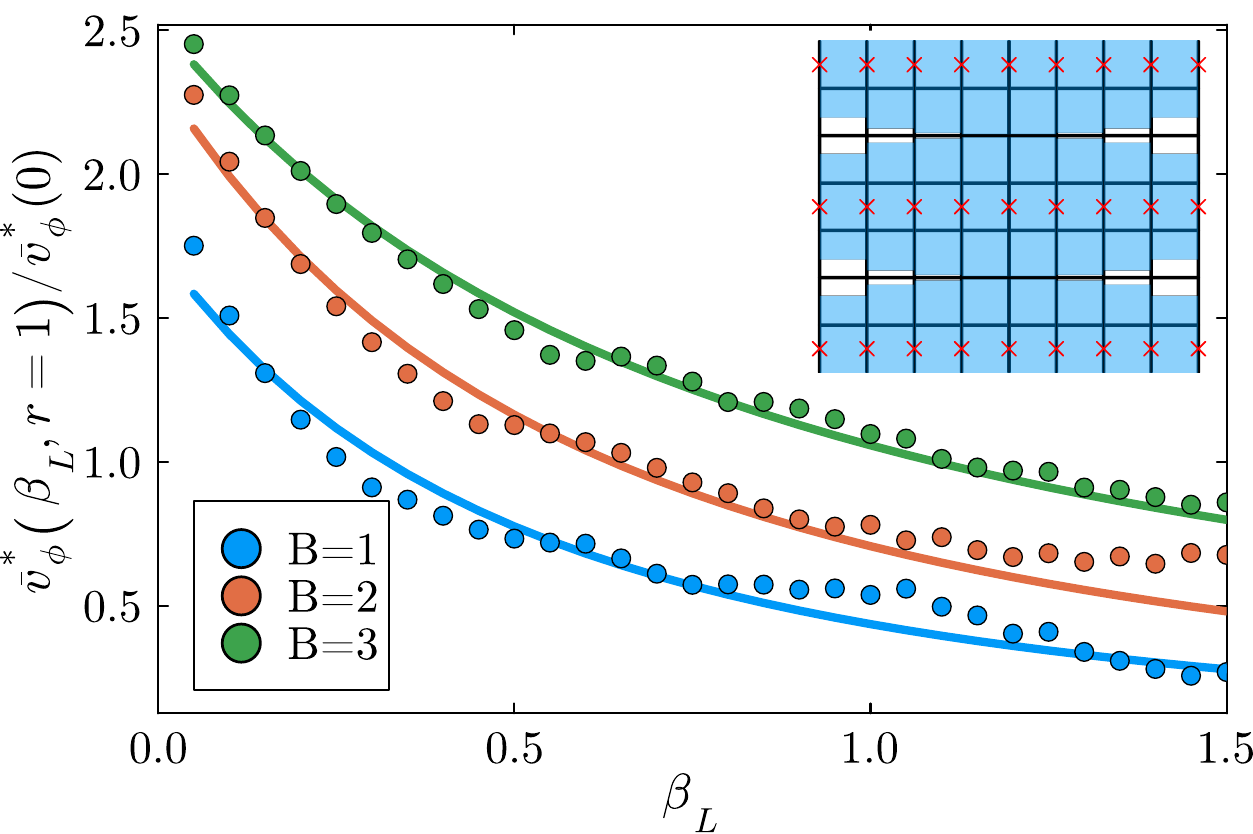}
    \includegraphics[width=0.475\linewidth]{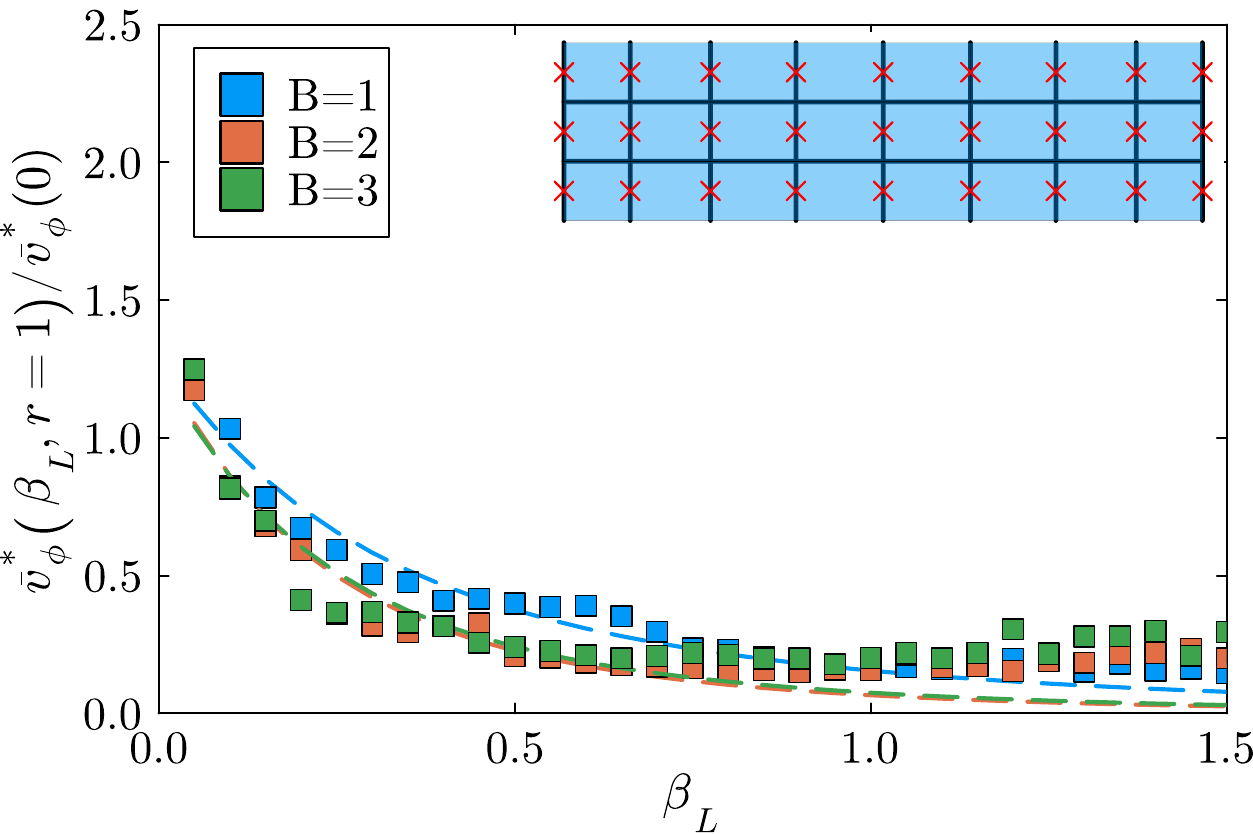}
    \caption{Maximum flux-to-voltage sensitivity of as a function of loop inductance for (left) mixed SQA  at \(r=1\) and \(N_p=16\), and (right) a conventional SQIF with physical area distributed the same way as the mixed SQA. The solid and dashed lines respectively mark power law fits to the numerical data points. We find that the performance of conventional SQIFs fall off much faster with the loop inductance when compared to mixed SQAs. Furthermore, increasing \(B\) actually improves the mixed SQA sensitivity, while it worsens the SQIF sensitivity.}
    \label{fig:sensitivity vs inductance}
\end{figure*}

While single SQUIDs exhibit \(\nu\approx 1\) \cite{Enpuku1995, Enpuku1995Method}, a \(N_p=16\) column array with \(r=1\) yields a mild decay exponent of \(\nu\approx1-2\) for mixed SQAs, compared to a much faster roll-off of \(\nu\approx 3-4\) in conventional SQIFs. Remarkably, increasing the number of bare loops \(B\) \emph{reduces} \(\nu\) for mixed SQAs, whereas increasing loop size in SQIFs \emph{increases} \(\nu\). In mixed SQAs, increasing the number of bare loops strengthens inductive screening and reduces the effective loop inductance experienced by the active SQUIDs. Conversely, in conventional SQIFs, physically enlarging the loop areas leads directly to higher parasitic inductances, yielding steeper sensitivity drop-offs. Consequently, mixed SQAs outperform conventional SQIFs by factors of \(2-4\times\) across realistic inductance regimes (see Fig.~\ref{fig:sensitivity vs inductance}). A detailed breakdown of scaling exponents is provided in Table~\ref{table: scaling exponents}.

\begin{table}[!ht]
\centering
\begin{tabular}{|c|c|c|c|} 
 \hline
    & \(N_p=4\) & \(N_p=8\) & \(N_p=16\) \\ [0.5ex] 
 \hline
 \(B=1\) & \(0.26\) & \(0.92\) & \(2.0\) \\ 
 \hline
 \(B=2\) & \(0.16\) & \(0.76\) & \(1.72\) \\
\hline
 \(B=3\) & \(0.12\) & \(0.67\) & \(1.27\) \\ [1ex] 
 \hline
\end{tabular}
\caption{Scaling exponents \(\nu\) of the maximum flux-to-voltage sensitivity calculated for mixed SQAs at \(r=1\) for varying number of bare loops \(B\), and number of columns in the array \(N_p\).}
\label{table: scaling exponents}
\end{table}

\section{Scaling with Size}\label{sec: scaling}

In this section, we study how the device characteristics scale with system size. First, let us understand the scaling in the limit of \(\b_L\rightarrow 0\). From Eq.\eqref{sensitivity ib=1 phi=0}, we see that \(\Bar{v}_{\phi}^{*}\propto B\cdot M\cdot N_p\) \textit{i.e.} the sensitivity scales with every dimension of the device. The peak-to-peak voltage, and the voltage range however, are expected to just scale with the number of rows, \(\Bar{v}^{*}, \Delta\Bar{v}\propto M\), while their dependence on the number of columns is more complicated. The linearity of the device on the other hand decreases as \(\mathrm{SFDR}\sim v_1/v_3 \propto (B\cdot N_p)^{-1}\). These trends are summarised in Table \ref{table:analytical scaling}.

\begin{table}[!ht]
\centering
\begin{tabular}{|c|c|c|c|} 
 \hline
 Characteristic Scaling & Rows & Columns & Bare Loops \\ [0.5ex] 
 \hline
 \(\Bar{v}_{\phi}^{*}\) & \(M\) & \(\approx N_p\) & \(\approx B\) \\ 
 \hline
 \(\Delta\Bar{v}\) & \(M\) & \(-\) & \(-\) \\
\hline
 \(\mathrm{SFDR}\) & \(1\) & \(\approx 1/N_p\) & \(\approx 1/B\) \\ [1ex] 
 \hline
\end{tabular}
\caption{Analytical scaling of the device characteristics in the limit of zero inductance. A \(-\) implies a non-trivial dependence with no simple scaling.}
\label{table:analytical scaling}
\end{table}

\begin{figure}[!ht]
    \centering
    \includegraphics[width=1.0\linewidth]{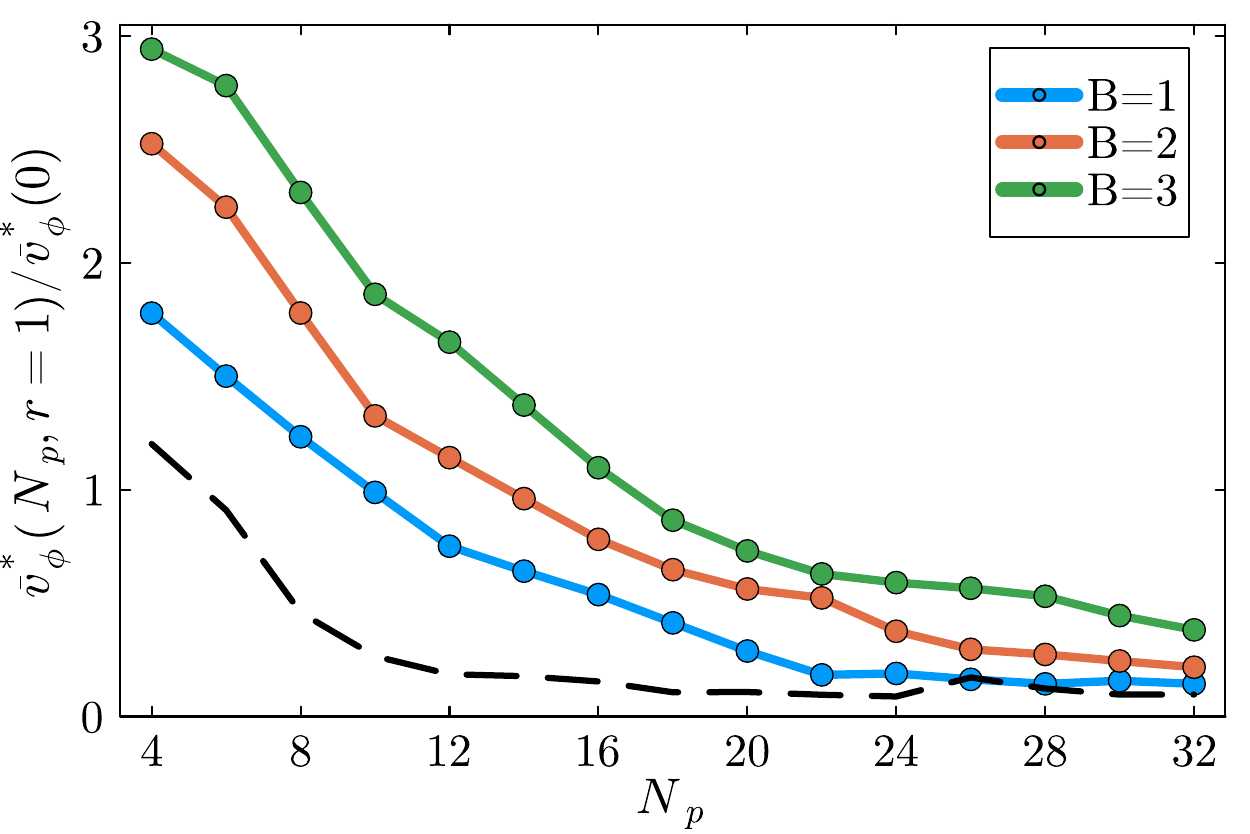}
    \caption{Normalised maximum flux-to-voltage sensitivity of a mixed SQA calculated at fixed \(r=1\) and \(\b_L=1\) as a function of the number of columns \(N_p\). Note that the normalisation factor depends quasi-linearly on \(N_p\) as \(\bar{v}_{\phi}^{*}(0) = M\pi\sqrt{N_p(N_p+2)/3}\). Consequently, the absolute value of the maximum possible sensitivity actually increases with \(N_p\) initially up-to \(N_p^{*}\in (12, 16)\) after which it starts decreasing due to inductive screening effects. The black dashed line marks the scaling of the sensitivity of a conventional SQIF with physical area spread.}
    \label{fig:scaling}
\end{figure}

However, at finite inductance, the scaling changes drastically as shown in Fig.~\ref{fig:scaling}. We find that the maximum sensitivity increases with \(N_p\) till about \(N_p\approx 12-20\), after which the sensitivity starts dropping \cite{mitchell_quantum_2019, kolotinskiy_choice_2021}. Crucially, this system-size performance degradation is significantly mitigated by increasing the number of bare loops \(B\). Furthermore, we again see that mixed SQAs consistently outperform conventional SQIFs as system sizes are increased by factors of up-to \(10\times\).

\section{Discussion}

In this work, we have presented a systematic study of mixed SQUID arrays (SQAs), providing key theoretical insights into how their performance compares to conventional SQIFs featuring physical loop-area dispersion. By incorporating junction-free (bare) superconducting loops into uniform SQUID topologies, we derived analytical expressions for the effective area spread and demonstrated that mutual inductive coupling drives the loop inductances toward a more uniform and reduced magnitude. Using a weak-coupling perturbation expansion ($\beta_L \to 0$), we obtained closed-form approximations for critical performance metrics, including flux-to-voltage sensitivity, linearity, and dynamic working range. Subsequently, we extended these findings to realistic inductance regimes ($\beta_L \sim \mo(1)$) via full numerical simulations. Crucially, our results reveal that mixed SQAs exhibit controlled and highly favourable scaling with both system size and inductance, successfully overcoming the severe performance degradation caused by inductive screening in traditional SQIF designs.\par

Beyond ideal RCSJ dynamics, several practical noise and physical implementation factors further highlight the advantages of mixed SQAs over conventional SQIFs. First, in real operational environments at \(4.2\:\mathrm{K}\) or \(77\:\mathrm{K}\), thermal fluctuations ($\Gamma = 2\pi k_B T / I_c \Phi_0$) cause phase rounding on junction I-V curves and damp the flux-to-voltage transfer function \cite{Ambegaokar_1969, tesche_dc_1977, koelle_high-transition-temperature_1999, enpuku1993effect}. Thermal noise degradation in SQUIDs is heavily accentuated by the screening parameter $\beta_L$ \cite{martinis_signal_1985, tesche_dc_1977}. Because mixed SQAs actively suppress the effective self-inductance ($\beta_{L,n}^{\mathrm{eff}} < \beta_L$) through inductive coupling to bare loops, the effective noise is lowered relative to conventional SQIFs. Consequently, mixed SQAs are inherently more resilient to thermal noise rounding and anti-peak smearing. Second, regarding fabrication yield, physical area-spread SQIFs are highly vulnerable to junction parameter variations ($\delta I_c / I_c$) across large junction networks \cite{kornev_design_2011, kornev_prospects}. mixed SQAs mitigate this failure mode structurally: by using uniform physical loop dimensions, the design requires significantly fewer active Josephson junctions (reduced by a factor of \(1/B\)), reducing the probability of encountering critical current outliers across the array \cite{GREENBERG2002156}.\par

Our comparative modelling also employs a conservative baseline for conventional SQIF performance. By assuming that loop self-inductances scale with perimeter as $\beta_{L,n}^{\mathrm{SQIF}} = \beta_L \sqrt{a_{n,\mathrm{eff}}/a_0}$ (characteristic of square loops or washer geometries) \cite{Drung2007highly, kornev_superconducting_2014}, we evaluate a best-case scenario for standard SQIFs. If a steeper linear scaling $\beta_{L,n} \propto a_{n,\mathrm{eff}}$ were assumed (e.g., elongated strip loops), standard SQIF sensitivity would drop off even faster with exponents $\nu \approx 4-6$, making our demonstrated $2-4\:\times$ to $10\:\times$ sensitivity enhancements a conservative lower bound. Additionally, while our model adopts nearest-neighbour inductive coupling as a tight-binding baseline, modern micro-fabrication methods routinely incorporate continuous superconducting ground planes or tightly coupled micro-strip layouts to confine magnetic flux return paths, validating nearest-neighbour coupling as a physically robust approximation \cite{dalichaouch_effects_2014, Tolpygo_self, Tolpygo_mutual, Tolpygo_2021}. \par

Our analysis also provides concrete design rules for scaling mixed SQAs up to large array dimensions. While conventional SQIFs hit a sharp performance ceiling at array widths of $N_p \approx 12-20$ columns due to runaway inductive screening, mixed SQAs maintain high voltage sensitivity across significantly larger dimensions. By tuning the number of bare loops per row $B$ and maintaining an inductive aspect ratio near $r \lesssim 1$, the size-dependent roll-off in flux sensitivity is dramatically softened. At realistic loop inductance values (\(\beta_L \sim 1\)), mixed SQAs yield sensitivity gains of \(2-4\:\times\) in moderate arrays and up to \(10\:\times\) in large-scale devices compared to equivalent SQIF layouts. This superior scaling behaviour highlights the promise of mixed SQAs for high-gain RF amplifiers and ultra-sensitive absolute magnetometers requiring high output power.\par

Finally, while our model accounts for primary mutual inductive coupling between adjacent bare and SQUID loops, future work could extend this framework to include long-range parasitic inductances, non-planar geometries, and spatial variations in bias currents. Experimental realisation and testing of these theoretical predictions using standard niobium \cite{Monaghan_synthetic_2025}, or high-$T_c$ YBCO thin-film fabrication processes (where controlling inductive screening is even more crucial in maintaining performance \cite{GREENBERG2002156, Cho_2019}) will be a crucial next step. Overall, the insights established in this work confirm and clearly demonstrates that mixed SQUID arrays resolve longstanding bottlenecks in SQUID array design, paving the way for a new class of highly scalable, high-performance quantum sensors.\par

\section{Acknowledgements}
A. B, J. L. M., and G. C. T are supported by the Australian Commonwealth Government through Research Training Program Scholarships. J. L. M. is also funded via the Ian Croser Research Scholarship in Engineering at Adelaide University. Giuseppe C. Tettamanzi acknowledges funding from the Australia’s Economic Accelerator Innovate Program scheme project number IV240100119. The concept of Synthetic Spread of the Areas and the subsequent idea of Mixed SQUID arrays are the subject of an international patent application no. PCT/AU2025/050991.

\bibliography{ref}

\appendix 
\onecolumngrid
\clearpage

\section{Circuit Modelling}\label{app: circuit modelling}
A generic \(N_s\times N_p\) mixed SQA with a mix of rows filled with \(N_p+1\) uniform Josephson junctions in parallel, and bare superconducting rows with no junctions is considered in this work as shown in Fig.~\ref{fig:device schematic} \cite{Monaghan_synthetic_2025}. The device has a total of \(N_F=N_s\cdot N_p\) superconducting loops out of which \(N_{sq}=M\cdot N_p\) are SQUIDs with a total of \(N_{jj}=M\cdot(N_p+1)\) junctions. The dynamic gauge-invariant phase of each junction is tracked in a vector, \(\bvf=(\vf_1\,,...\,,\vf_{N_{jj}})^{T}\), where the index of the vector is a composite index comprised of the row and column \(\vf_i\equiv \vf_{(m, n)}\) \(m \in \{1\,,...\,,M\}\) and \(n\in\{1\,,...\,,N_p+1\}\). Each loop has a corresponding loop current : \(\bj_{S}=(j_{S}^{1}\,,...\,,j_{S}^{N_{sq}})^{T}\) in the SQUID loops, and \(\bj_{B}=(j_B^1\,,...\,,j_B^{N_b})^{T}\) (\(N_b = N_F-N_{sq}\)) in the bare loops. \par

Now, the branch currents through each junction \(\bi=(i_1\,,...\,,i_{N_{JJ}})^{T}\) may be written in terms of the applied bias current \(i_b\) and loop currents as
\begin{equation}
    \label{current relation}
    \bi = \bK_{b}i_b + K\cdot\bj_{S}\,,
\end{equation}
where \(\bj_B\) cannot affect the currents through the junction by definition. \(K\) is the \(N_{jj}\times N_{sq}\) incidence matrix which relates the junction branches to SQUID loops \textit{s.t.} the entries \(K_{ij} = \pm 1\) if the \(i^{th}\) junction branch lies on the \(j^{th}\) SQUID loop (with the sign determined by the relative orientation of the loop with the branch), or zero otherwise. Similarly, \(\bK_b\) is the bias current incidence vector which can be treated by fixing a path in the device (the most straightforward being a straight line connecting the bias source to the drain) and assuming a current \(i_b\) flowing on that chosen path \cite{chua_linear_1987, garcia_mona_2021}. \par 

Next, let us consider flux quantization, which on the bare superconducting loops takes the form
\begin{equation}
    \label{bare loop flux quantization}
    \boldsymbol{0} = 2\pi\boldsymbol{\phi}_B-\b_L\left(L_C^{T}\cdot\bj_{S}+L_{B}\cdot\bj_{B}+\boldsymbol{L}_{b}^{B} i_b\right)\,,
\end{equation}
with \(\boldsymbol{\phi}_B=\phi (1\,,...\,,1)^{T}\). \(L_B\) is the symmetric normalised inductance matrix of all the bare loops, while \(L_C\) is the off-diagonal inductance coupling the bare-loop currents to the fluxes in the SQUID loops. \(\bL_b^B\) is the inductive contribution of the applied bias current to the bare loops which depends on the bias injection and the path basis chosen from bias source to drain. \(\b_L=2\pi I_c L_{\circ}/\Phi_0\) is the overall inductance normalisation factor where \(L_{\circ} = 2(L_h+L_v)\) is the total loop inductance of a SQUID with two horizontal and two vertical wire segments. Now, since \(L_B\) is square symmetric, its inverse exists and hence the bare loop flux quantization allows us to solve for \(\bj_B\). Meanwhile on the actual SQUID loops, flux quantization takes the form
\begin{equation}
    \label{SQUID loops flux quantization}
    K^T\cdot\bvf = 2\pi\boldsymbol{\phi}_{S}-\b_L\left(L_{S}\cdot\bj_{S}+L_C\cdot\bj_{B}+\boldsymbol{L}_{b}^{S}i_b\right)\,,
\end{equation}
where \(K^T\cdot\bvf\) returns the phase difference of the Josephson junctions on each SQUID loop~\cite{Monaghan_synthetic_2025}. \(L_S\) is the symmetric normalized inductance matrix of the SQUID loops, while \(\bL_b^S\) is the inductive contribution of the bias current to the SQUID loops. Combining Eq.\eqref{bare loop flux quantization} with Eq.\eqref{SQUID loops flux quantization}, we find that the bare loops modify the effective fluxes, inductances, and bias current distributions in the device. Lastly, using Eq.\eqref{current relation}, we get \cite{Mitchell_2022, Monaghan_synthetic_2025}
\begin{equation}
    \label{flux quantization effective gauge invariant}
    \boldsymbol{i} = -\b_L^{-1} K\cdot L_{\mathrm{eff}}^{-1}\cdot\left(K^T\cdot\bvf-2\pi\boldsymbol{\phi}_{\mathrm{eff}}\right)+\bK_{b}^{\mathrm{eff}}i_b\,,
\end{equation}
where the effective parameters are as given in Eq.\eqref{effective parameters}.

\section{Effective Circuit Parameters}\label{app: circuit parameters}

We consider a \(N_s\times N_p\) SQUID array with \(N_s=(M+1)\cdot B + M\) and \(N_p\in2\mathbb{Z}\). We have \(N_{jj}=M\cdot(N_p+1)\) Josephson junctions forming \(N_{sq}=M\cdot N_p\) SQUIDs and \(N_b=(M+1)\cdot B\cdot N_p\) bare loops. The junctions, SQUID loops, and bare loops are all numbered in a row-major (first count the columns in each row then change the row) way.\par 

By construction, both \(K\) and \(D\) should only depend on the column and not the row. Hence, we get the following simple structure in these matrices
\begin{equation}
    \label{D and K matrix}
    K = \mathds{1}_{M}\otimes U_{N_p}^T(1, -1)\,,
\end{equation}
where \(U_{N_p}(a, b)\) is a upper bi-diagonal Toeplitz matrix of the form
\begin{equation}
    \label{B matrix definition}
    [U_{N_p}(a, b)]_{N_p\times (N_p+1)} = \begin{pmatrix}
        a & b & 0 & 0 & \hdots & 0\\
        0 & a & b & 0 &\hdots & 0 \\
        0 & 0 & a & b & \hdots & \vdots\\
        \vdots & \vdots & \ddots & \ddots  & \ddots & \\
        0 & \hdots &  & 0 & a & b
    \end{pmatrix}\,.
\end{equation}
Now let us formulate the inductance matrices. We assume that the horizontal wires of the device have a partial inductance \(L_h\), while the vertical segments have partial inductance \(L_v\). Consequently, we have that
\begin{equation}
\label{L_sq form}
    L_{S}=\frac{\b_L}{2}\mathds{1}_M\otimes\left[\frac{1}{(1+r)} T_{N_p}(1, 1) + 2\frac{r}{(1+r)} \mathds{1}_{N_p}\right]\,,\:\:\:r=\frac{L_h}{L_v}\,,
\end{equation}
and \(T_n(a, b)\) is the generalized tri-diagonal \(n\times n\) Toeplitz matrix with \(2a\) on the diagonals and \(-b\) on the off-diagonals as 
\begin{equation}
    \label{TN form}
    [T_n(a, b)]_{n\times n} = \begin{pmatrix}
        2a & -b & 0 & \hdots & 0\\
        -b & 2a & -b & \hdots & 0\\
        0 & -b & 2a & \ddots & \vdots \\
        \vdots & & \ddots &\ddots & -b\\
        0 & \hdots & & -b & 2a
    \end{pmatrix}\,.
\end{equation}
 The term proportional to \(L_v\) i.e. the vertical wire segment inductances, couples each SQUID loop to the inductance of the two wire segments in the same loop, as well as the wire-segments in the horizontally neighboring loops (with the opposite chirality). Similarly, the term proportional to \(L_h\) couples SQUID loops with the two horizontal segments in the same loop as well. Next, the bare inductances have a similar form  
\begin{equation}
    \label{bare inductance matrix}
    L_B = \frac{\b_L}{2}\mathds{1}_{M+1}\otimes\left[\frac{1}{(1+r)}\mathds{1}_{B}\otimes T_{N_p}(1, 1) + \frac{r}{(1+r)}T_B(1, 1)\otimes\mathds{1}_{N_p}\right]\,.
\end{equation}
Lastly, the coupling inductance has the form 
\begin{equation}
    \label{coupling inductance form}
    L_C = -\frac{\b_{L}}{2}\frac{r}{(1+r)}\left[U_M(1, 0)\otimes A_1 + U_M(0, 1)\otimes A_2\right]\otimes \mathds{1}_{N_p}\,,
\end{equation}
where \([A_1]_{1\times B}\) and \([A_2]_{1\times B}\) such that \(A_1^{ij}=\d_{i, 1}\d_{j, B}\), while \(A_2^{ij}=\d_{i, 1}\d_{j, 1}\). The term involving \(A_1\) is the coupling between the SQUID row to the bare row above it, while \(A_2\) is the coupling to the bare row below.\par

Integrating out the bare loops change the effective characteristics of the device as seen in Eq.\eqref{flux quantization effective gauge invariant}. The key quantity required to evaluate the renormalized parameters in Eq.~\eqref{effective parameters} is \(L_B^{-1}\). To that end, let us define \(L_B^{-1}=\b_L^{-1}(1+r)[\mathds{1}_{M+1}\otimes G(r)]\) where \(G(r) \equiv \left[\mathds{1}_B\otimes T_{N_p}(1, 1) + r T_B(1, 1)\otimes\mathds{1}_{N_p}\right]^{-1}\). We need to evaluate the form of \(G\) which may be done in the spectral representation of \(G^{-1}(r) = H(r) = \left[\mathds{1}_B\otimes T_{N_p}(1, 1) + r T_B(1, 1)\right]\). \(H(r)\) can be interpreted as the discretized equivalent of the Hamiltonian for a particle in an infinite potential well in 2-d with \(B\times N_p\) sites. The kinetic energy is just the 2-d dispersion for a particle with hopping \(1\) along \(\hat{y}\), and hopping \(r\) along \(\hat{x}\), and a chemical potential \(\mu = 2+2r\). Hence,
\begin{equation}
    \label{2d dispersion}
    \e(\bk) = 2(1+r)-2r\cos(k_x)-2\cos(k_y)\,,\:\:\:\bk=(k_x, k_y) = \pi\left(\frac{i_x}{B+1}, \frac{i_y}{N_p+1}\right)\,,\:\:\:i_x\in\{1, ..., B\}\,,i_y\in\{1, ..., N_p\}\,.
\end{equation}
The corresponding eigenfunctions have the following form
\begin{equation}
    \label{2d eigenfunctions}
    \braket{x, y}{\psi_{\bk}} = \sqrt{\frac{2}{B+1}}\sqrt{\frac{2}{N_p+1}}\sin(k_x x)\sin(k_y y)\,.
\end{equation}
Note that \(\e(\bk)>0 \:\forall \:\bk\) and hence the Hamiltonian is strictly positive-definite. Hence, the inverse is well-defined in the spectral representation, and can be written as
\begin{equation}
    \label{form of G}
    G(r) = \sum_{\bk}\frac{\ket{\psi_{\bk}}\bra{\psi_{\bk}}}{\e(\bk)}\,,
\end{equation}
with matrix elements
\begin{equation}
    \label{matrix inverse form of G}
    [G(r)]_{b_1, n_1}^{b_2, n_2} = \frac{4}{(B+1)(N_p+1)}\sum_{\bk}\frac{\sin(k_x b_1)\sin(k_x b_2)\sin(k_y n_1)\sin(k_y n_2)}{2(1+r)-2\cos(k_y)-2r\cos(k_x)}\,,\,.
\end{equation}

\subsection{Effective Fluxes}

\begin{figure}[!ht]
    \centering
    \includegraphics[width=0.475\linewidth]{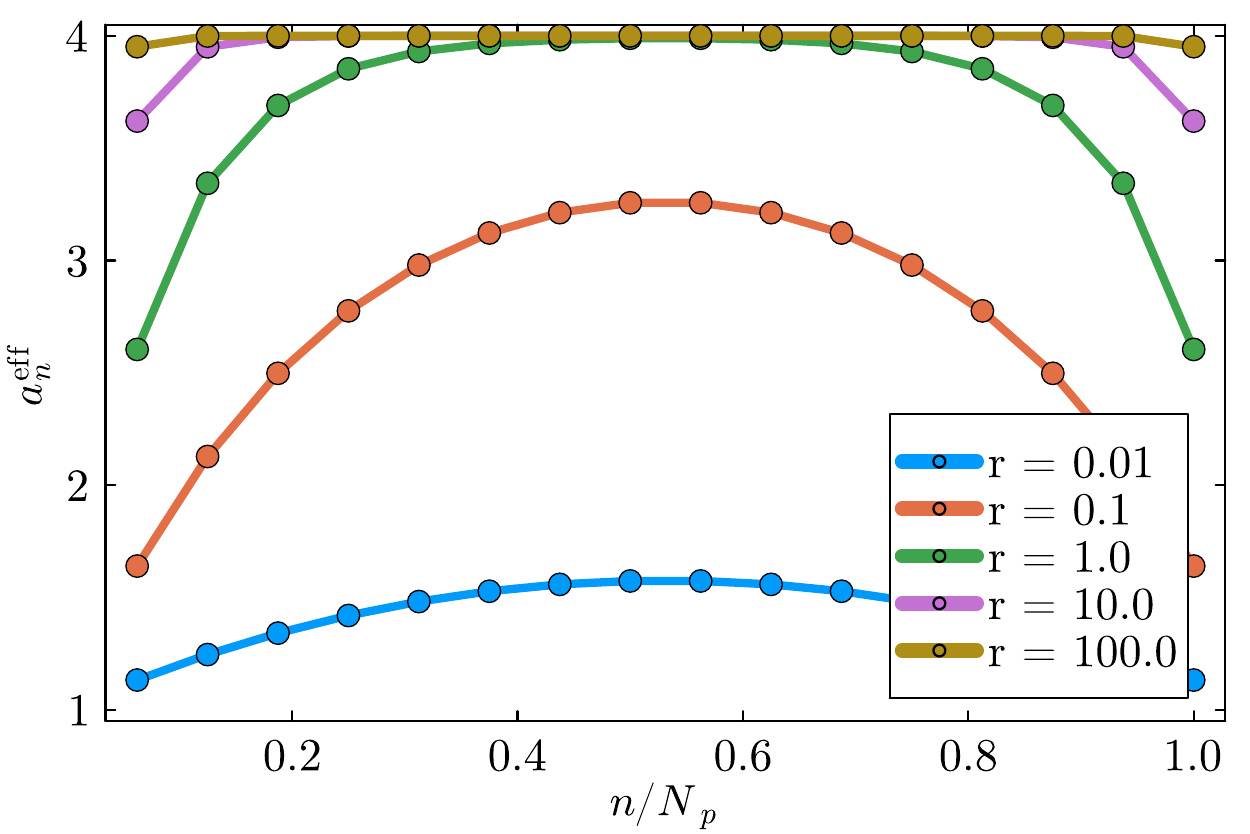}
    \includegraphics[width=0.475\linewidth]{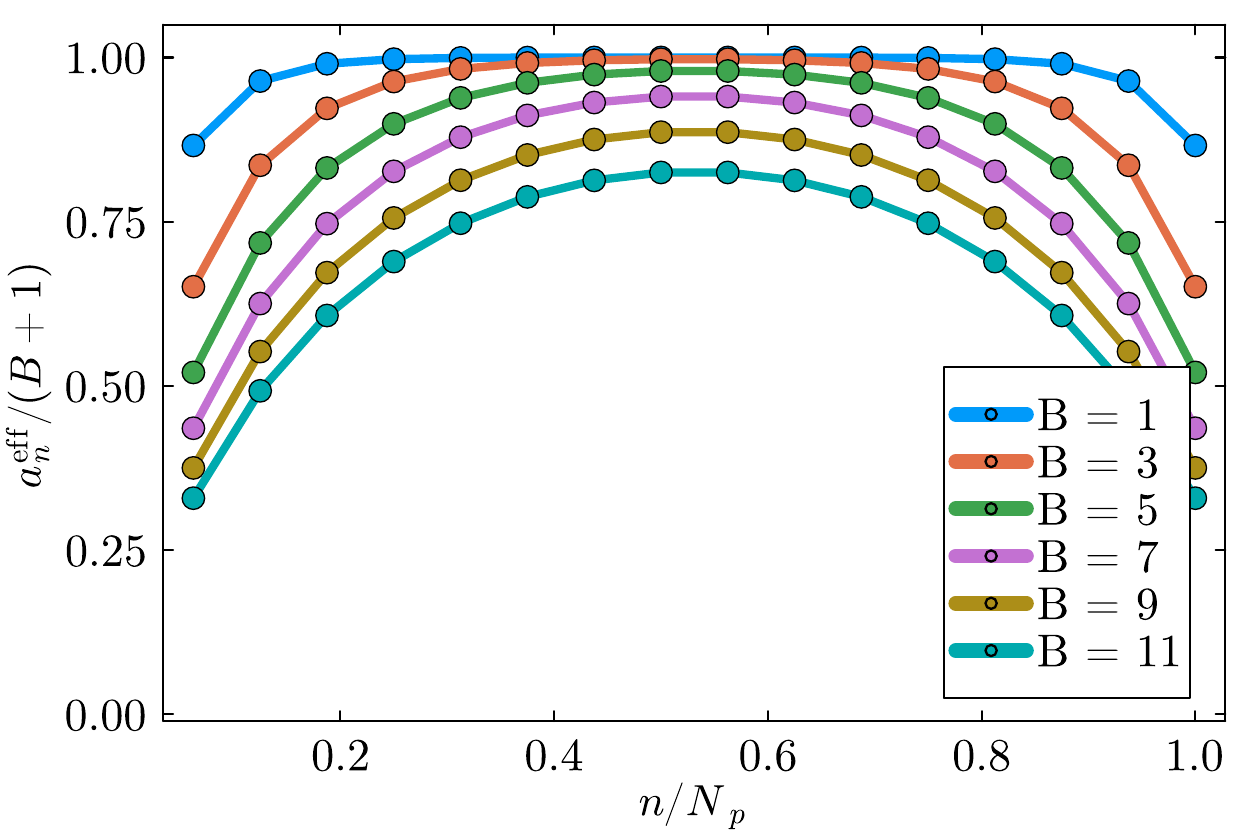}
    \caption{Effective area distribution \(a_{n}^{\eff}(r)\) after integrating out the bare loops in a device with \(N_p=16\). (a) Fixed \(B=3\) and varying \(r\). (b) Fixed \(r=1\) and varying \(B\).}
    \label{fig:Effective area distribution vs r and b}
\end{figure}

The effective fluxes take the form
\begin{equation}
    \label{effective fluxes with G}
    \boldsymbol{\phi}_{\eff} =\boldsymbol{\phi}_{S}+r\left\{B_M(1, 0)\otimes \left[(A_1 \otimes \mathds{1}_{N_p})\cdot G\right] + B_M(0, 1)\otimes \left[(A_2 \otimes\mathds{1}_{N_p})\cdot G\right]\right\}\cdot\boldsymbol{\phi}_B\,.
\end{equation}
Now, assuming that the entire array is exposed to uniform external flux \(\phi\), we have that \(\phi_S = \phi(1, ..., 1)^{T}_{N_{sq}}\) and \(\phi_B = \phi(1, ..., 1)^{T}_{N_b}\). Defining \(\boldsymbol{\phi}_{\eff}=\phi \boldsymbol{a}_{\eff}\), we get that
\begin{equation}
    \label{effective area}
    \begin{split}
      a_{n}^{\eff}(r) &= 1 + \sum_{b_2=1}^{B}\sum_{n_2=1}^{N_p}\left([G(r)]_{1, n}^{b_2, n_2} + [G(r)]_{B, n}^{b_2, n_2}\right)\\
      &= 1+\frac{8r}{(B+1)(N_p+1)}\sum_{\bk}\frac{\sin(k_x)\sin(k_y n)}{2(1+r)-2\cos(k_y)-2r\cos(k_x)}\left(\sum_{b_2=1}^{B}\sin(k_x b_2)\right)\left(\sum_{n_2=1}^{N_p}\sin(k_y n_2)\right)\,.
    \end{split}
\end{equation}
Next, we use the identity for \(\sum_{j=1}^{n}\sin(j\theta) = \sin((N_p+1)\theta/2)\sin(N_p\theta/2)/\sin(\theta/2)\) along with the form of \(k_x = i_x\pi/(B+1)\) to get
\begin{equation}
    \label{an eff step 2}
    \begin{split}
      a_{n}^{\eff}(r) &= 1+\frac{4r}{(B+1)}\sum_{i_x\in \mathrm{Odd}}^{B}\cos^2\left(\frac{k_x}{2}\right)\left[\sum_{n_2=1}^{N_p}\frac{2}{(N_p+1)}\sum_{k_y}\frac{\sin(k_y n)\sin(k_y n_2)}{2\cosh(\theta_{i_x})-2\cos(k_y)}\right]\,,
    \end{split}
\end{equation}
where \(\cosh(\theta_{i_x}) = 1+2r\sin^2(k_x/2)\). The term in the square bracket is the Green's function of a 1d particle in a box at \(\mu=2\cosh(\theta_{i_x})\). The form of this is known exactly \cite{tan2018explicitinversetridiagonalmatrix} and can be substituted back into Eq.\eqref{an eff step 2}
\begin{equation}
    \label{2d Greens function step 2}
    a_{n}^{\eff}(r) = 1+\frac{4r}{(B+1)}\sum_{i_x\in \mathrm{Odd}}^{B}\cos^2\left(\frac{k_x}{2}\right)\left[\sum_{n_2=1}^{N_p}\frac{\sinh(\min(n,n_2)\theta_{i_x})\sinh((N_p+1-\max(n, n_2))\theta_{i_x})}{\sinh(\theta_{i_x})\sinh((N_p+1)\theta_{i_x})}\right]\,.
\end{equation}
This sum can then be simplified to give 
\begin{equation}
    \label{1d Greens function}
    \sum_{n_2=1}^{N_p}\frac{\sinh(\min(n,n_2)\theta_{i_x})\sinh((N_p+1-\max(n, n_2))\theta_{i_x})}{\sinh(\theta_{i_x})\sinh((N_p+1)\theta_{i_x})} = \frac{1}{2(\cosh(\theta_{i_x})-1)}\left\{1-\frac{\cosh\left(\left(\frac{(N_p+1)}{2}-n\right)\theta_{i_x}\right)}{\cosh\left(\frac{(N_p+1)}{2}\theta_{i_x}\right)}\right\}\,.
\end{equation}
Hence, we end up with
\begin{equation}
    \label{2d Greens function step 3}
    a_{n}^{\eff}(r) = 1+\frac{2}{(B+1)}\sum_{i_x\in \mathrm{Odd}}^{B}\cot^2\left(\frac{k_x}{2}\right)\left[1-\frac{\cosh\left(\left(\frac{(N_p+1)}{2}-n\right)\theta_{i_x}\right)}{\cosh\left(\frac{(N_p+1)}{2}\theta_{i_x}\right)}\right]\,.
\end{equation}
Lastly, we can use the identity \(\sum_{i_x\in\mathrm{Odd}}^{B}\cot^2(i_x\pi/2(B+1)) = B(B+1)/2\) to get back Eq.\eqref{effective areas s=1 b>1} with \(\theta_{b}(r) = \cosh^{-1}\left(1+2r \sin^2\left(\frac{b \pi}{2(B+1)}\right)\right)\)

\subsection{Effective Inductances}\label{app: effective inductance}

\begin{figure}[!ht]
    \centering
    \includegraphics[width=0.55\linewidth]{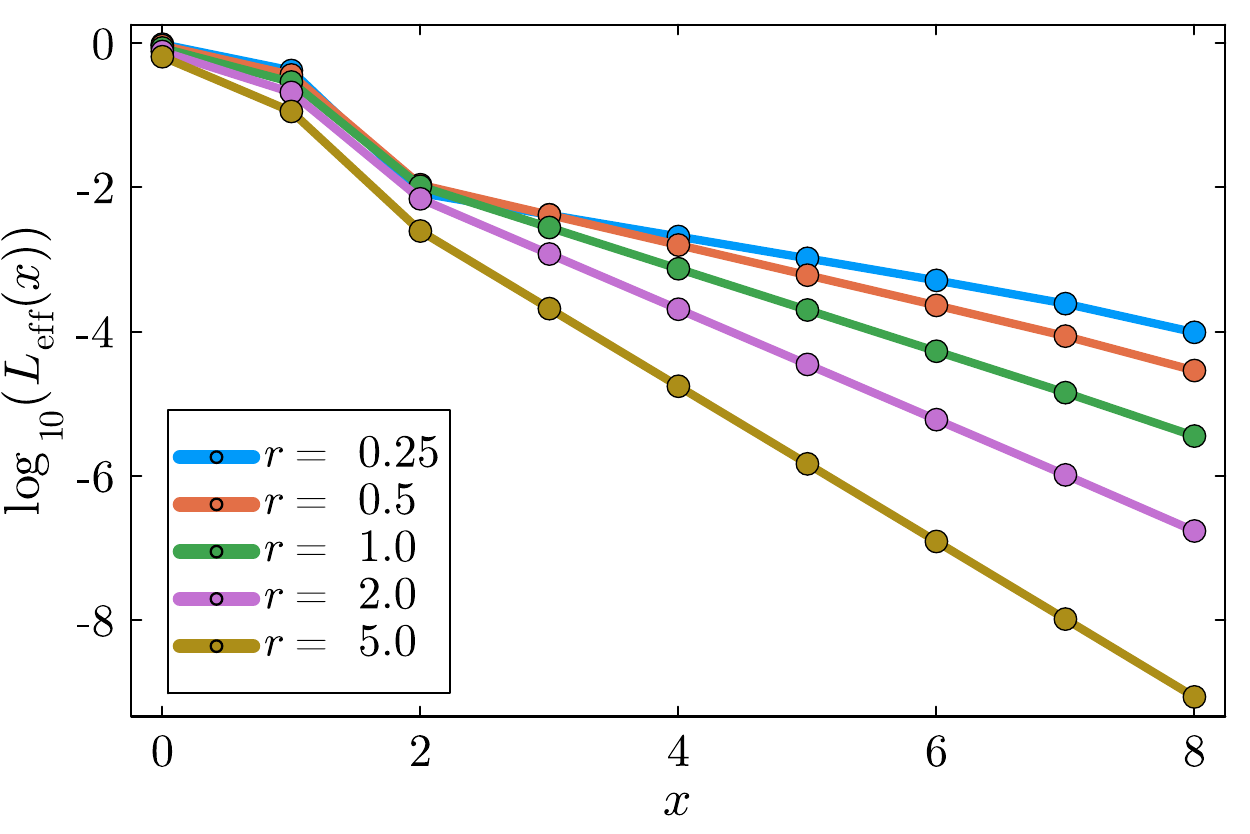}
    \caption{The normalized mutual inductance strength (log-scale) as a function of distance between SQUID loops in the same row for different values of \(r\). We find an exponentially decaying long-range behavior.}
    \label{fig:Mutual inductance}
\end{figure}

Let us now shift our attention to calculating the renormalized inductance matrix \(L_{\eff}\) as given in Eq.\eqref{effective parameters}. We will focus on the simpler case where \(B=1\) but the procedure can be generalized to \(B>1\) as well. For \(B=1\), \(L_B\) has the simplified structure of a tridiagonal matrix \(L_B = \b_L/(1+r)\mathds{1}_{M+1}\otimes T_{N_p}(1+r, 1)\). Hence, \(L_B^{-1} = \b_L^{-1}(1+r)[\mathds{1}_{M+1}\otimes G_1(r)]\) where \(G_1(r)\) is the 1d Green's function as used and elucidated in Eq.\eqref{2d Greens function step 2} and Eq.\eqref{1d Greens function}. Explicitly, we have
\begin{equation}
    \label{G1}
    [G_1(r)]_{ij} = \frac{\sinh(\min(i, j)\theta)\sinh((N_p+1-\max(i, j))\theta)}{\sinh(\theta)\sinh((N_p+1)\theta)}\,,\:\:\:\cosh(\theta)=1+r\,.
\end{equation}
The coupling inductance \(L_C\) also simplifies to \(L_C = -\b_{L}(r/(1+r))[B_M(1,1)\otimes\mathds{1}_{N_p}]\). Hence, we get
\begin{equation}
    \label{Leff b=1}
    \begin{split}
     L_{\eff} &= \b_{L, v}\left(\mathds{1}_{M}\otimes T_{N_p}(1+r, 1) - r^2 (B_M(1, 1)\cdot B_M^T(1, 1))\otimes G_1(r)\right)  \\
     & = \b_{L, v}\left(\mathds{1}_{M}\otimes T_{N_p}(1+r, 1) - r^2 T_M(1, -1)\otimes G_1(r)\right)\,.
    \end{split}
\end{equation}
The diagonal elements of Eq.\eqref{Leff b=1} have the form mentioned in Eq.\eqref{effective self-inductance}. The effective inductance matrix also has mutual inductance terms arising due to coupling to the bare loops. In Fig.~\ref{fig:Mutual inductance}, we plot the mutual inductance as a function of distance between SQUID loops.

\section{Setting up the Perturbation Series}\label{app: perturbation series}

Let us go back to the master equation given in Eq.\eqref{flux quantization effective gauge invariant}
\begin{equation}
    \label{master equation}
    -\b_L^{-1}K\cdot L_{\mathrm{eff}}^{-1}\cdot\left(K^T\cdot\bvf-2\pi\boldsymbol{\phi}_{\mathrm{eff}}\right)+\bK_{b}^{\mathrm{eff}}i_b = \boldsymbol{i} = \sin(\bvf) + \frac{\doe \bvf}{\doe \tau} + \b_c \frac{\doe^2 \bvf}{\doe \tau^2}\,.
\end{equation}
Now we notice that the LHS of this equation \emph{only} depends on difference-phases in each row, and is invariant under an overall global shift of all phases (in each row of the device independently). Mathematically, this implies that there exist vectors \(\bx_i\) \textit{s.t.}
\begin{equation}
    \label{null vectors}
    K^T\cdot\bx_i = 0\,,\:\:\:[\bx_i]_{(m, n)} = \d_{i, m}\,,\:\:\:{i, m}\in\{1, ..., M\}\,,n\in\{1, ..., N_p+1\}\,,
\end{equation}
which are null vectors of the matrix \(D\). As a result, we can split all the junction phases in \(\bvf\) as
\begin{equation}
    \label{phase parametrization 1}
    \bvf = \sum_{i=1}^{M}\bx_i \theta_i + P\cdot \delta\bvf\equiv W\cdot\begin{pmatrix}
        \bth\\
        \d\bvf
    \end{pmatrix}\,,\:\:\:W = \begin{pmatrix}
        X & P
    \end{pmatrix}\,,\:\:\:X = \begin{pmatrix}
        \bx_1 & \bx_2 & \hdots & \bx_M
    \end{pmatrix}\,.
\end{equation}
where \(\theta_i\) is the phase of the central junction in each SQUID row, and \([\d\bvf]_{M\cdot N_p}\) are the difference-phases in each row. \([P]_{M\cdot (N_p+1) \times M\cdot N_p}\) is a projection matrix \textit{s.t.} \(K^T\cdot P = \mathds{1}_{M}\otimes\mathds{1}_{N_p}\). There is a freedom in the choice of \(P\), depending on how we define \(\theta\). If we define \(\theta_i\) as the average phase of the \(i^{th}\) row, \(\theta_i = \sum_{n=1}^{(N_p+1)}\varphi_{i, n}/(N_p+1)\), then \(P\) is fixed to be \(P = K\cdot (K^T\cdot K)^{-1}\). As a result, we get that \(K^T\cdot\bvf = K^T\cdot P\cdot\d\bvf = \d\bvf\) which makes explicit
the fact that the LHS of Eq.\eqref{master equation} does not depend on the global sum-phases \(\theta_i\).\par

Next, we wish to do a series expansion of the phase in powers of the loop inductance \(\b_L\) as
\begin{equation}
    \label{series expansion}
    \begin{split}
        \theta_i &= \theta_i^0 + \b_L\theta_i^1 + ...\,,\\
        \delta \bvf &= \delta\bvf_0 + \b_L \delta\bvf_1 + ...\,.
    \end{split}
\end{equation}
Assuming \(\b_c\ll 1\) to avoid the hysteretic regime, we have
\begin{equation}
    \label{master equation rescaled}
    -K\cdot L_{\mathrm{eff}}^{-1}\cdot\left(\delta\bvf-2\pi\boldsymbol{\phi}_{\mathrm{eff}}\right)+\b_L\bK_{b}^{\mathrm{eff}}i_b = \b_L\sin(\bvf) + \b_L\frac{\doe \bvf}{\doe \tau}\,.
\end{equation}
Now, we want to split the equation into the dynamics of the sum-phases and difference-phases, which we can do using the projection matrix \(P\) and the null-vectors \(\bx_i\). Remember that \(K = D^{T}\) and hence, \(\bx_i^{T}\cdot K = 0\) and \(P^T\cdot K = \mathds{1}_M\otimes\mathds{1}_{N_p}\). 

We get the following
\begin{equation}
    \label{master equation decoupled}
    \begin{split}
        -L_{\mathrm{eff}}^{-1}\cdot\left(\delta\bvf-2\pi\boldsymbol{\phi}_{\mathrm{eff}}\right)+\b_L P^T\cdot\bK_{b}^{\mathrm{eff}}i_b & = \b_L P^T\cdot\sin(\bvf) + \b_L P^T\cdot\frac{\doe \bvf}{\doe \tau}\,,\\
        \bx_i^T\cdot\bK_{b}^{\mathrm{eff}}i_b & = \bx_i^T\cdot\sin(\bvf) + \bx_i^T\cdot\frac{\doe \bvf}{\doe \tau}\,.
    \end{split}
\end{equation}
We can simplify this further. Using the definitions in Eq.\eqref{effective parameters}, we see that \(P^T\cdot\bK_{\eff}^b = P^T\cdot\bK_{b}-L_{\eff}^{-1}\cdot\bL_{b}^{\eff}\), and \(\bx_i^{T}\cdot\bK_{\eff}^b = \bx_i^{T}\cdot\bK_b = (N_p+1)\). Furthermore, using \(\bx_i^{T}\cdot\bx_j = (N_p+1)\d_{ij}\), we get a matrix ODE and a system of scalar ODEs as
\begin{equation}
    \label{master equation decoupled scalar+vector}
    \begin{split}
         \b_L \sum_{i=1}^{M}(P^T\cdot \bx_i)\frac{\doe \theta_i}{\doe \tau} + \b_L (P^T\cdot P)\cdot\frac{\doe \delta\bvf}{\doe \tau} & = -\b_L P^T\cdot\sin(\bvf) -L_{\mathrm{eff}}^{-1}\cdot\left(\delta\bvf-2\pi\boldsymbol{\phi}_{\mathrm{eff}}\right)+\b_L i_b ( P^T\cdot\bK_b-L_{\mathrm{eff}}^{-1}\cdot\bL_{b}^{\eff}) \,,\\
       (N_p+1)\frac{\doe \theta_i}{\doe \tau} +  (P^T\cdot \bx_i)^T \cdot \frac{\doe \delta\bvf}{\doe \tau} & = - \bx_i^T\cdot\sin(\bvf) + i_b (N_p+1) \,.
    \end{split}
\end{equation}
Now note that \(P^T\cdot\bx_i = (K^T\cdot K)^{-1}\cdot K^T\cdot\bx_i = 0\), and \(P^T\cdot P = (K^T\cdot K)^{-1}\). This gives us that \((P^T\cdot P)^{-1}\cdot P^T = K^T\) and hence, we end up with the equations in Eq.\eqref{master equation decoupled scalar+vector final}. 

\subsection{Leading Order}\label{app: leading order}
\begin{equation}
    \label{zeroth order equations}
    \begin{split}
        \d\bvf_0 & = 2\pi\boldsymbol{\phi}_{\eff} \,,\\
        \frac{\doe \theta_m^0}{\doe \tau}  & = i_b - \frac{\bx_m^T}{(N_p+1)}\cdot\sin(\bvf_0) \,.
    \end{split}
\end{equation}
Consequently, the phase of \emph{every} junction in a filled row is determined by effectively a single dynamic average-phase, and the applied flux \(\phi\) as

To leading order in \(\b_L\), we have that
\begin{equation}
    \label{zeroth order equations 1}
    \begin{split}
        \d\bvf_0-2\pi\boldsymbol{\phi}_{\eff} & = 0\,,\\
        \frac{\doe \theta_i^0}{\doe \tau}  & = i_b - \frac{\bx_i^T}{(N_p+1)}\cdot\sin(\bvf_0) \,.
    \end{split}
\end{equation}
Using the first equation, and substituting back into the second equation, we get
\begin{equation}
    \label{zeroth order equations for theta}
    \frac{\doe \theta_i^0}{\doe \tau} \equiv v_i^0(\tau) = i_b - i_c(\phi)\sin(\theta_i^0) \,,
\end{equation}
where we used 
\begin{equation}
    \label{critical current matrix form}
    \bx_i^{T}\cdot\sin(\bvf_0) = (N_p+1)i_c(\phi)\sin(\theta_i^0)
\end{equation}
with \(i_c(\phi)\) being the same as defined in Eq.\eqref{modified critical current}. The solution to this equation, assuming that \(\theta_i^0(\tau=0)=0\) has the form
\begin{itemize}
    \item \textbf{Over-biased : \(i_b>i_c(\phi)\)}:
    \begin{equation}
        \label{zeroth order solution for theta}
        \begin{split}
             \sin(\theta_i^0(\t)) &= \frac{i_c(\phi)-i_b\cos(\w_0 \t+\g)}{i_b-i_c(\phi)\cos(\w_0 \t+\g)}\,,\\
             \cos(\theta_i^0(\t)) &= \frac{\w_0\sin(\w_0 \t+\g)}{i_b-i_c(\phi)\cos(\w_0\t+\g)}\,,\\
             v_{i}^{0}(\t) &= \frac{\w_0^2}{i_b-i_c(\phi)\cos(\w_0\t+\g)}\,,
        \end{split}
    \end{equation}
    where
    \begin{equation}
        \label{zeroth order voltage 1}
        \w_0 = \Bar{v}_0 = \bar{v}_0(\phi, i_b) =  \sqrt{i_b^2-i_c^2(\phi)}\,,\:\:\:\cos(\g(\phi))=\frac{i_c(\phi)}{i_b}\,.
    \end{equation}
    \item \textbf{Under-biased : \(i_b<i_c(\phi)\)}:
    \begin{equation}
        \label{zeroth order solution for theta underbiased}
        \begin{split}
             \sin(\theta_i^0(\t)) &= \frac{i_b\cosh(\w_0 \t+\g)-i_c(\phi)}{i_c(\phi)\cosh(\w_0 \t+\g)-i_b}\xrightarrow{\w_0\tau\gg 1}\frac{i_b}{i_c(\phi)}\,,\\
             \cos(\theta_i^0(\t)) &= \frac{\w_0\sinh(\w_0 \t+\g)}{i_c(\phi)\cosh(\w_0\t+\g)-i_b}\xrightarrow{\w_0\tau\gg 1}\frac{\w_0}{i_c(\phi)}\,,\\
             v_{i}^{0}(\t) &= \frac{\w_0^2}{i_c(\phi)\cosh(\w_0\t+\g)-i_b}\xrightarrow{\w_0\tau\gg 1}\frac{2\w_0^2}{i_c(\phi)}e^{-\w_0\t}\rightarrow 0\,,
        \end{split}
    \end{equation}
    where
    \begin{equation}
        \label{zeroth order voltage underbiased}
        \w_0 =  \sqrt{i_c^2(\phi)-i_b^2}\,,\:\:\:\bar{v}_0=0\,,\:\:\:\cosh(\g(\phi))=\frac{i_c(\phi)}{i_b}\,.
    \end{equation}
\end{itemize}
The total voltage over the full device is uniform
\begin{equation}
    \label{zeroth order voltage}
    \bv_0(\t) = v_0(\t) \bI\,.
\end{equation}

The DC time averages of a function \(f(t)\) is defined as
\begin{equation}
    \label{time average}
    \overline{f}\equiv \lim_{T\rightarrow\infty}\frac{1}{T}\int_0^{T}dt f(t) = \frac{1}{T_f}\int_0^{T_f}dt f(t)\,,
\end{equation}
where \(T_f\) is a single period if \(f\) is a periodic function. When the device is over-biased, the voltage is periodic and hence its DC average is the same as its average over a single time-period \(T_0=2\pi/\w_0\). This time-average will amount to \(\theta_0(T_0)/T_0\neq 0\) since when the device is over-biased, the phase is linearly increasing with small oscillation around the linear trend.\par

\subsubsection{Linearized Equations in the under-biased regime}\label{app: linearized}
For the under-biased case, the phases are exponentially damped and settle into steady-state value as \(\w_0\t\rightarrow\infty\). As a result, the voltage vanishes since it is the time-derivative of the phase. In fact, in the fully under-biased regime, one can set all time-derivatives to zero in the long-time limit of Eq.\eqref{master equation decoupled scalar+vector}. Hence, we get a system of non-linear \emph{static} equations as
\begin{equation}
    \label{master equation underbiased}
    \begin{split}
         \b_L P^T\cdot\sin(\bvf^*) +\hat{L}_{\mathrm{eff}}^{-1}\cdot\left(\delta\bvf^*-2\pi\boldsymbol{\phi}_{\mathrm{eff}}\right)+ & = \b_L i_b ( P^T\cdot\bK_b-\hat{L}_{\mathrm{eff}}^{-1}\cdot\hat{\bL}_{b}^{\eff}) \,,\\
       \bx_i^T\cdot\sin(\bvf^*) & =  i_b (N_p+1) \,,
    \end{split}
\end{equation}
where \(\bvf^*\) is the long-time steady-state limit of the junction phases in the under-biased regime. Now there are two ways one can tackle this non-linear equation. Either we can continue the perturbation series, just in the simplified steady-state regime (as we will do in the next section), or we can linearize the equation for small \(i_b\) and solve the linear system of equations given by
\begin{equation}
    \label{master equation underbiased linearized}
    \begin{split}
         \b_L P^T\cdot\bvf^* +\hat{L}_{\mathrm{eff}}^{-1}\cdot\left(\delta\bvf^*-2\pi\boldsymbol{\phi}_{\mathrm{eff}}\right)+ & = \b_L i_b ( P^T\cdot\bK_b-\hat{L}_{\mathrm{eff}}^{-1}\cdot\hat{\bL}_{b}^{\eff}) \,,\\
       \bx_i^T\cdot\bvf^* & =  i_b (N_p+1) \,.
    \end{split}
\end{equation}
The second equation can be solved to get
\begin{equation}
    \label{solution for theta star underbiased}
    \theta_i^* = i_b - \frac{\bx_i^T\cdot P \cdot \d \bvf^*}{(N_p+1)}\,,
\end{equation}
which can be substituted back into the first equation to get (at zero external flux)
\begin{equation}
    \label{underbiased linearized solution}
    \bvf^{*} = i_b\bI + \b_L i_b(\mathds{1}-M)\cdot P \cdot \left[\mathds{1} + \b_L \hat{L}_{\eff}\cdot P^T\cdot(\mathds{1}-M)\cdot P\right]^{-1}\cdot \hat{L}_{\eff}\left(P^T\cdot\bK_b - P^T\cdot\bI - \hat{L}_{\eff}^{-1}\cdot\bL_b^{\eff}\right)\,,
\end{equation}
where \(M = \sum_i \bx_i \bx_i^T/(N_p+1)\). Since we are in the linearized regime, \(\bi^*\approx\bvf^*\) and hence this gives us the bias current distribution in the under-biased regime \(i_b\ll1\) for arbitrary inductance \(\b_L\).

\begin{figure}
    \centering
    \includegraphics[width=0.475\linewidth]{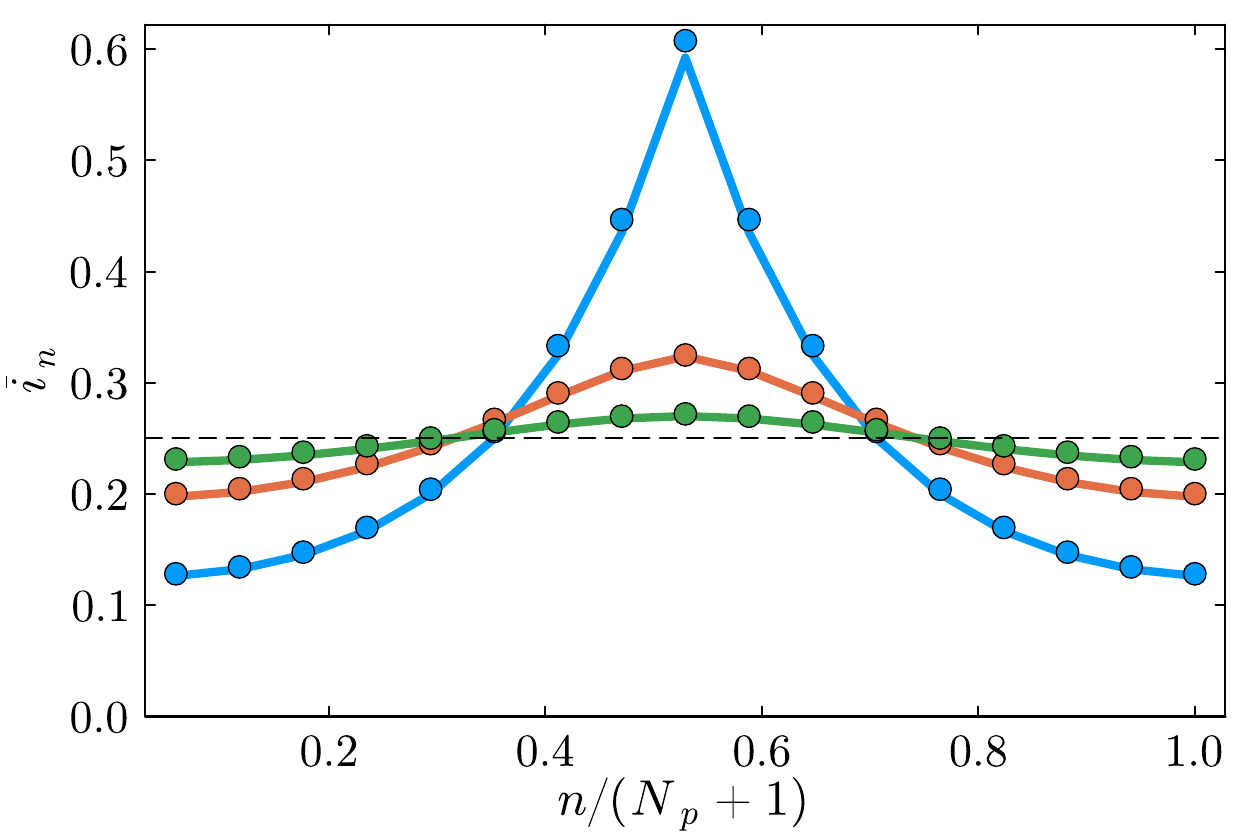}
    \includegraphics[width=0.475\linewidth]{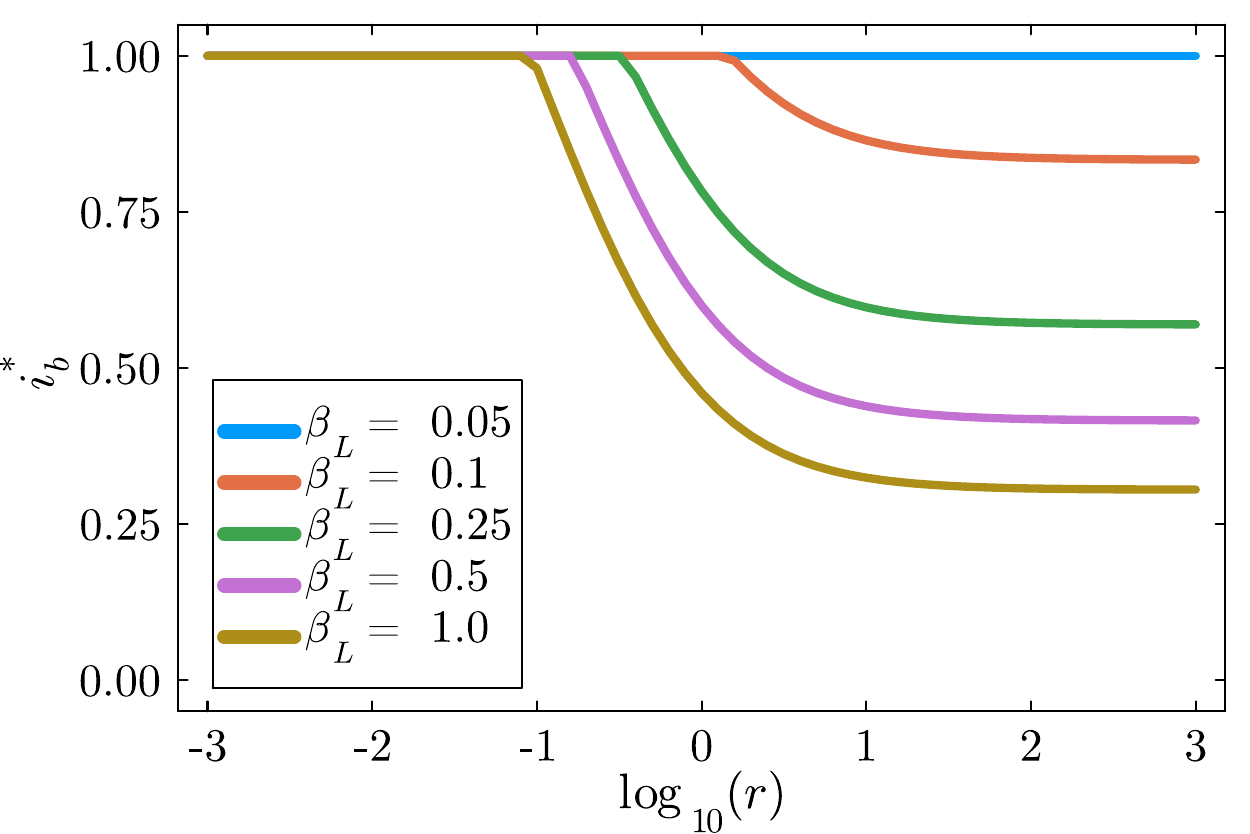}
    \caption{(Left) The bias current distribution in the Off-state of the SQUID array at a bias \(i_b=0.25\) for the different rows in the device : first row at the edge (blue), second row (orange), and third row (green). As one goes into the bulk, the current distribution becomes more and more uniform (black dashed line). (Right) The device critical current \(i_b^{*}=\min_{i_b}(\max \bvf^{*}(i_b) \geq \pi/2)\) which marks the transition of the array into the intermediate mixed-state as a function of \(r\) for different values of the inductance. }
    \label{fig:bias current distribution appendix}
\end{figure}

\subsection{Sub-leading order}\label{app: sub-leading}
To next order in \(\b_L\), we get corrections as
\begin{equation}
    \label{first order equation 1}
    \begin{split}
         \frac{\doe \delta\bvf_0}{\doe \tau} & = - K^T\cdot\sin(\bvf_0) -(K^T\cdot K)\cdot\hat{L}_{\mathrm{eff}}^{-1}\cdot\delta\bvf_1+i_b ( K^T\cdot\bK_b-(K^T\cdot K)\cdot\hat{L}_{\mathrm{eff}}^{-1}\cdot\hat{\bL}_b^{\eff}) \,,\\
       \frac{\doe \theta_i^1}{\doe \tau}  & = - \bx_i^T\cdot C_0(\t)\cdot \bvf_1 \,,
    \end{split}
\end{equation}
where \(C_0(\t)\) is a diagonal matrix whose diagonal entries are given by \(\cos(\bvf_0(\t))\). The first equation does not involve any time-derivative of \(\bvf_1\) and hence can be solved immediately as
\begin{equation}
    \label{first order solution dphi}
    \delta\bvf_1 = i_b ( \hat{L}_{\eff}\cdot P^T\cdot\bK_b-\hat{\bL}_b^{\eff})-\hat{L}_{\eff}\cdot P^{T}\cdot\sin(\bvf_0) \,,\:\:\: P^T = (K^T\cdot K)^{-1}\cdot K^T\,.
\end{equation}
The time-derivative of \(\bvf_1\) has the form
\begin{equation}
    \label{first order solution dphi/dt}
    \begin{split}
       \frac{\doe \delta\bvf_1}{\doe \t} &= -v_0(\t) \hat{L}_{\eff}\cdot P^{T}\cdot \cos(\bvf_0(\t)) \,,\\
       \frac{\doe^2 \delta\bvf_1}{\doe \t^2} &= -\hat{L}_{\eff}\cdot P^{T}\cdot \left[ \frac{\doe v_0(\t)}{\doe\t} \cos(\bvf_0(\t)) -v_0^2(\t)\sin(\bvf_0(\t))\right]\,.
    \end{split}
\end{equation}

Now note that \(\delta\bvf_1(\t)\) given in Eq.\eqref{first order solution dphi} is \emph{purely periodic in \(\t\)} with period given by the zeroth-order voltage \(T_0=2\pi/\w_0\) when the device is over-biased. Hence,
\begin{equation}
    \label{first order correction to voltage dphi1}
    \overline{\frac{\doe\bvf_1}{\doe\tau}} = \frac{\Bar{v}_0}{2\pi}[\delta\bvf_1(\t)]_0^{T_0} = 0\,.
\end{equation}
When the device is under-biased, the difference-phase \(\delta\bvf_1(\t)\) becomes a constant after an exponentially fast evolution. Hence, its time-derivative vanishes after it has settled, and therefore the DC average of the derivative is zero again. Therefore, the sub-leading correction to the difference phase will not contribute to the DC averaged voltage of the device.\par 

Next, we study the corrections to the uniform phase \(\theta_m^1\). We have that
\begin{equation}
    \label{first order equation theta}
   \frac{\doe\theta_i^1}{\doe\tau} = -\frac{1}{(N_p+1)}\sum_{j=1}^{M}\left(\bx_i^{T}\cdot C_0(\t)\cdot\bx_j\right)\theta_j^1-\frac{1}{(N_p+1)}\bx_i^{T}\cdot C_0(\t)\cdot P\cdot\delta\bvf_1 \,.
\end{equation}
We can simplify the first term on the RHS as
\begin{equation}
    \label{theta1 term}
    \begin{split}
       \left(\bx_i^{T}\cdot C_0(\t)\cdot\bx_j\right) & = \d_{ij}\sum_{n=1}^{N_p+1}\cos(\bvf_0^{(i, n)})\\
       & = \d_{ij}\frac{\doe}{\doe\theta_i^0}\sum_{n=1}^{N_p+1}\sin(\bvf_0^{(m, n)}) = \frac{\doe}{\doe\theta_i^0}[(N_p+1)i_c(\phi)\sin(\theta_i^0)]\\
       & = \d_{ij}(N_p+1)i_c(\phi)\cos(\theta_i^0)\,.
    \end{split}
\end{equation}
Now note that \(v_0(\theta_0(\t)) = i_b-i_c(\phi)\sin(\theta_0(\t))\), and hence \(\doe v_0/\doe\tau = -i_c(\phi)\cos(\theta_0)v_0\). Therefore, we get that
\begin{equation}
    \label{first order equation theta 1}
    (N_p+1)v_0(\t)\frac{\doe}{\doe\tau}\left(\frac{\theta_i^1}{v_0}\right) = -\bx_i^{T}\cdot C_0(\t)\cdot P\cdot\delta\bvf_1 \,.
\end{equation}
To solve this, we express the RHS as forces \(F_i^1\) with the following expression
\begin{equation}
    \label{force term}
    \begin{split}
        F_i^1(\tau)\equiv -& \bx_i^{T}\cdot C_0(\t)\cdot P\cdot\delta\bvf_1\\
        = & -i_b\left[\bx_i^{T}\cdot C_0(\t)\cdot(\hat{L}_{\eff}^P\cdot\bK_b-P\cdot\hat{\bL}_b^{\eff})\right]+\left[\bx_i^{T}\cdot C_0(\t)\cdot\hat{L}_{\eff}^{P}\cdot\sin(\bvf_0)\right]\,,
    \end{split}
\end{equation}
where \(\hat{L}_{\eff}^{P} = P\cdot\hat{L}_{\eff}\cdot P^T\). Before we continue, it is useful to split \(\sin(\bvf_0)\), \(\cos(\bvf_0)\), and \(C_0(\t)\) into components which consists of a time-dependent scalar function of \(\theta_0(\t)\), and a \emph{time-independent} vector or matrix. We have that
\begin{equation}
    \label{splitting vectors}
    \begin{split}
        \sin(\bvf_0) &= \sin(\theta_0(\t))\cos(2\pi P\cdot\bph_{\eff})+\cos(\theta_0(\t))\sin(2\pi P\cdot\bph_{\eff})\,,\\
        \cos(\bvf_0) &= \cos(\theta_0(\t))\cos(2\pi P\cdot\bph_{\eff})-\sin(\theta_0(\t))\sin(2\pi P\cdot\bph_{\eff})\,,\\
        C_0(\t) &= \cos(\theta_0(\t))C_0(2\pi P\cdot\bph_{\eff})-\sin(\theta_0(\t))S_0(2\pi P\cdot\bph_{\eff})\,,
    \end{split}
\end{equation}
where \(C_0\) and \(S_0\) are diagonal matrices whose diagonal entries are given by \(\cos(2\pi P\cdot\bph_{\eff})\) and \(\sin(2\pi P\cdot\bph_{\eff})\) respectively. Substituting Eq.\eqref{splitting vectors} back into Eq.\eqref{force term}, we get
\begin{equation}
    \label{Fi time dependent split}
    \begin{split}
        F_i^1(\theta_0(\t)) &= \sin(\theta_0(\t))f_s^i + \cos(\theta_0(\t))f_c^i + \cos(\theta_0(\t))\sin(\theta_0(\t))f_{cs}^i + \sin^2(\theta_0(\t))f_{ss}^i+\cos^2(\theta_0(\t))f_{cc}^i\,,
    \end{split}
\end{equation}
where
\begin{equation}
    \label{Fi coefficients}
    \begin{split}
        f_s^i &= i_b\bx_i^{T}\cdot S_0(2\pi P\cdot\bph_{\eff})\cdot(\hat{L}_{\eff}^P\cdot\bK_b-P\cdot\hat{\bL}_b^{\eff})\,,\\
        f_c^i &= -i_b\bx_i^{T}\cdot C_0(2\pi P\cdot\bph_{\eff})\cdot(\hat{L}_{\eff}^P\cdot\bK_b-P\cdot\hat{\bL}_b^{\eff})\,,\\
        f_{cs}^i &= -\bx_i^{T}\cdot S_0(2\pi P\cdot\bph_{\eff})\cdot\hat{L}_{\eff}^{P}\cdot\sin(2\pi P\cdot\bph_{\eff})+\bx_i^{T}\cdot C_0(2\pi P\cdot\bph_{\eff})\cdot\hat{L}_{\eff}^{P}\cdot\cos(2\pi P\cdot\bph_{\eff})\,,\\
        f_{ss}^i &= -\bx_i^{T}\cdot S_0(2\pi P\cdot\bph_{\eff})\cdot\hat{L}_{\eff}^{P}\cdot\cos(2\pi P\cdot\bph_{\eff})\,,\\
        f_{cc}^i &= \bx_i^{T}\cdot C_0(2\pi P\cdot\bph_{\eff})\cdot\hat{L}_{\eff}^{P}\cdot\sin(2\pi P\cdot\bph_{\eff})\,.
    \end{split}
\end{equation}
We can utilize \emph{mirror} symmetry in our device to get rid of some of these coefficients. Remember that \(\phi_{\eff}\) is symmetric when reflected along the central column. However, \(P\) is anti-symmetric and hence the difference-phase \(P\cdot\phi_{eff}\) is anti-symmetric \textit{w.r.t} the central column. Since \(\sin(-x) = -\sin(x)\), we get that \(\sin(P\cdot\phi_{\eff})\) and \(S_0(P\cdot\phi_{\eff})\) are both anti-symmetric under such a mirror symmetry. However, note that \(\bx_i, \hat{L}_{\eff}^{P},\) and \(\bK_b\) are all symmetric under the same mirror symmetry. Now, any overlap between an anti-symmetric and a symmetric vector would vanish. We see that
\begin{equation}
    \label{vanishing coefficients}
    f_s^i = f_{ss}^i = f_{cc}^i = 0 \:\forall \: i\,.
\end{equation}

Next, Let us define 
\begin{equation}
    G_i^1(\t) = \int_0^{\t}d\t' \frac{F_i^1(\t')}{(N_p+1)v_0(\t')}
\end{equation}
Then, we get that
\begin{equation}
    \label{theta 1 solution}
    \theta_i^1(\t) = v_0(\t)G_i^1(\t)\,,\:\:\:\frac{\doe\theta_i^1}{\doe\t} = \frac{\doe v_i^0}{\doe\t}G_i^1(\t)+\frac{1}{(N_p+1)}F_i^1(\t)\,.
\end{equation}
Our job boils down to calculating the time-integral of \(f(\t)/v_0(\t)\) where \(f(\t) = \cos(\theta_0(\t))\,,\) and \(\cos(\theta_0(\t))\sin(\theta_0(\t))\,,\). We can make use of the fact that
\begin{equation}
    \label{voltage derivative}
    i_c(\phi)\cos(\theta_0(\t)) = -\frac{1}{v_0(\t)}\frac{\doe v_0}{\doe \t}\,.
\end{equation}
This allows us to get closed form for all the desired integrals as follows (Remember that if \(\theta_0(0) = 0\) then \(v_0(0) = i_b\) by Eq.\eqref{zeroth order equations for theta})
\begin{subequations}
    \label{time integrals}
    \begin{align}
        \int_{0}^{\t}d\t'\frac{\cos(\theta_0(\t'))}{v_0(\t')} & = -\frac{1}{i_c(\phi)} \int_{v_0(0)}^{v_0(\t)}d v_0 \frac{1}{v_0^2} = \frac{1}{i_c(\phi)}\left[\frac{1}{v_0}\right]_{i_b}^{v_0(\t)} = \frac{1}{i_c(\phi)}\left[\frac{1}{v_0(\t)}-\frac{1}{i_b}\right]\,,\\
        \int_{0}^{\t}d\t'\frac{\sin(2\theta_0(\t'))}{2v_0(\t')} & = -\frac{1}{i_c^2(\phi)} \int_{v_0(0)}^{v_0(\t)}d v_0 \frac{i_b - v_0}{v_0^2} = \frac{1}{i_c^2(\phi)}\left[\frac{i_b}{v_0}+\ln(v_0)\right]_{i_b}^{v_0(\t)} = \frac{1}{i_c^2(\phi)}\left[\frac{i_b}{v_0(\t)}-1-\ln\left(\frac{i_b}{v_0(\t)}\right)\right]\,.
    \end{align}
\end{subequations}
Note that special care has to be taken when \(i_c(\phi)\rightarrow 0\) and the limiting behavior has to be extracted carefully. We see there are only two sources of time-dependence in \(G_i^1(\t)\) : \(1/v_0(\t)\) and \(\ln(v_0(\t))\). Collecting all the terms we get
\begin{equation}
    \label{theta1 expression final}
    \theta_i^1(\t) = \frac{1}{(N_p+1)}\left\{\left[\frac{f_c^i}{i_c(\phi)} + \frac{i_b f_{cs}^i}{i_c^2(\phi)}\right]-v_0(\t)\left[\frac{f_c^i}{i_b i_c(\phi)}+\frac{f_{cs}^i}{i_c^2(\phi)}\right] - v_0(\t)\ln\left(\frac{i_b}{v_0(\t)}\right)\left[\frac{f_{cs}^i}{i_c^2(\phi)}\right]\right\}\,.
\end{equation}

\begin{equation}
    \label{dtheta1/dt expression final}
    \begin{split}
       \frac{\doe\theta_i^1}{\doe\t} &= \frac{1}{(N_p+1)}\frac{\doe v_0}{\doe \t}\left\{-\left[\frac{f_c^i}{i_b i_c(\phi)}\right] -\ln\left(\frac{i_b}{v_0(\t)}\right)\left[\frac{f_{cs}^i}{i_c^2(\phi)}\right]\right\}\,,\\ 
       \frac{\doe^2\theta_i^1}{\doe\t^2} &= \frac{1}{(N_p+1)}\left\{-\frac{\doe^2 v_0}{\doe \t^2}\left[\frac{f_c^i}{i_b i_c(\phi)}\right]-\left(\frac{\doe^2 v_0}{\doe \t^2}\ln\left(\frac{i_b}{v_0(\t)}\right)-\frac{1}{v_0(\t)}\frac{\doe v_0}{\doe \t}\right)\left[\frac{f_{cs}^i}{i_c^2(\phi)}\right]\right\}\,.
    \end{split}
\end{equation}
We find that \(\theta_1(\t)\) is also \emph{purely} periodic, and hence cannot contribute to the DC voltage. The total periodic correction to the voltage takes the form
\begin{equation}
    \label{voltage first order}
    \begin{split}
        \bv_1(\t) &= \sum_i \bx_i \frac{\doe\theta_i^1}{\doe\t} + P\frac{\doe\d\bvf_1}{\doe\t}\,.
    \end{split}
\end{equation}

\end{document}